\documentclass[aps,prl,twocolumn,showpacs,preprintnumbers]{revtex4}
\usepackage{amsmath}
\usepackage{dcolumn}
\usepackage{bm}
\usepackage{subfigure}
\usepackage{amsfonts}
\usepackage{amssymb}
\usepackage{makeidx}
\usepackage{epsfig}
\usepackage{graphicx}

\begin{document}

\title{\textbf{Variational formulation\ of the electromagnetic
radiation-reaction problem }}
\author{M. Tessarotto\thanks{%
Electronic-mail: M.Tessarotto@cmfd.univ.trieste.it}$^{a,b}$, C. Cremaschini$%
^{c}$ and M. Dorigo$^{d}$}
\affiliation{$^{a}$Department of Mathematics and Informatics, University of Trieste, Italy%
\\
$^{b}$Consortium of Magneto-fluid-dynamics, University of Trieste, Italy\\
$^{c}$International School for Advanced Studies (SISSA), Trieste, Italy\\
$^{d}$Department of Physics, University of Trieste, Italy}
\date{\today }

\begin{abstract}
A fundamental issue in classical electrodynamics is represented by the
search of the \textit{exact equation of motion for a classical charged
particle} under the action of its electromagnetic (EM) self-field - the
so-called \textit{radiation-reaction equation of motion (RR equation).} In
the past, several attempts have been made assuming that the particle
electric charge is localized point-wise (point-charge). These involve the
search of possible so-called "regularization" approaches able to deal with
the intrinsic divergences characterizing point-particle descriptions in
classical electrodynamics. In this paper we intend to propose a new solution
to this problem based on the adoption of a variational approach and the
treatment of finite-size spherical-shell charges. The approach is based on
three key elements: 1) the adoption of the relativistic synchronous Hamilton
variational principle recently pointed out (Tessarotto et al, 2006); 2) the
variational treatment of the EM self-field, for finite-size charges, taking
into account the exact particle dynamics; 3) the adoption of the axioms of
classical mechanics and electrodynamics. The new RR\ equation proposed in
this paper, departing significantly from previous approaches, exhibits
several interesting properties. In particular: a) unlike the LAD
(Lorentz-Abraham-Dirac) equation, it recovers a second-order ordinary
differential equation which is fully consistent with the law of inertia,
Newton principle of determinacy and Einstein causality principle and b)
unlike the LL (Landau-Lifschitz) equation, it holds also in the case of
sudden forces. In addition, it is found that the new equation recovers the
customary LAD equation in a suitable asymptotic approximation.
\end{abstract}

\maketitle

%\bmulticol

\section{1 - Introduction}

The goal of this paper is to investigate a well-known theoretical issue of
classical electrodynamics. This is concerned with the solution of the
so-called radiation-reaction problem (\emph{RR problem}), i.e., the
description of the dynamics of classical charges (charged particles) in the
presence of their EM self-fields. For contemporary science the possible
solution of the RR problem represents not merely an unsolved intellectual
challenge, but a fundamental prerequisite for the proper formulation of all
physical theories which are based on the description of relativistic
dynamics of classical charged particles. These involve, for example, the
consistent formulation of the relativistic kinetic theory of charged
particles and of the related fluid descriptions (i.e., the relativistic
magnetohydrodynamic equations obtained by means of suitable closure
conditions), both essential in plasma physics and astrophysics.

Surprisingly, until recently \cite{Tessarotto2008c} (hereon denoted Ref.A)
the problem has remained substantially unsolved, despite efforts spent by
the scientific community in more than one century of intensive theoretical
research (see related discussion in Ref.\cite{Dorigo2008a}; for a review see
Refs.\cite{Rohrlich1965}). In particular, still missing is the \emph{exact
relativistic equation of motion for a classical charged particle in the
presence of its electromagnetic (EM) self-field, }also known as\emph{\ }(%
\emph{exact})\emph{\ RR equation}. \ For definiteness, in the following we
shall consider the RR problem in the case of a flat (Minkowski) space-time,
although a similar problem can be posed, in principle, also for curved
space-time and in the context of a general-relativistic formulation. This
requires that $g\equiv \det \left\{ g_{\mu \nu }\right\} =-1,$ $g_{\mu \nu }$
denoting the Minkowski metric tensor with signature $(1,-1,-1,-1)$.

The equation, to be achieved exclusively in the framework of a
classical-mechanics description, should result \emph{non-asymptotic}\textit{.%
} Namely, the exact RR equation should not rely on any asymptotic expansion
(i.e., a truncated perturbative expansion), in particular for the
electromagnetic field generated by the charged particle, to be \ performed
in terms of any possible infinitesimal parameter which may characterize the
particle itself (assuming that in some sense the particle has a finite
"size", i.e., it is not point-like). On the other hand, by assumption, a
\emph{classical particle} should satisfy at least two basic properties: a)
to have no "internal structure" and b) to be spherically symmetric (when
seen with respect to the particle rest-frame).\ These hypotheses (which are
manifestly satisfied by point-particles), should be fulfilled also by
finite-size particles in which the mass and/or the electric charge have a
finite-size distribution. Hence, these parameters should (only) be related
to the radii of the mass and/or charge distributions. Following the
prescription pointed out in Ref.A, here we intend to prove that\ such an
equation can be obtained explicitly, without introducing any so-called
"regularization" scheme, i.e., leaving unchanged the axioms of classical
electrodynamics. The result is reached by considering classical \emph{%
finite-size charges, }and, more precisely,\emph{\ finite-size
spherical-shell charges }(in analogy to the classical\ Lorentzian model \cite%
{Lorentz}). For these particles the charge is considered spatially
distributed in a bounded 3D domain (i.e., characterized by a
finite-size charge distribution). In detail, the charge density -
when seen with respect to each particle rest-frame -\ is taken by
assumption: a) \emph{spherically symmetric}, b) \emph{radially
localized on a spherical surface }$\delta \Omega _{\sigma }$
having a finite radius $\sigma >0;$ c) \emph{quasi-rigid}, i.e.,
to remain constant on\emph{\ }$\delta \Omega _{\sigma }$ with
respect to the same reference frame. Hypotheses a)-c) are
manifestly all consistent with the above requirements for a
classical particle. Instead, as far as the mass distribution is
concerned, it is
assumed as point-wise localized in the center of the spherical surface $%
\delta \Omega _{\sigma }.$ This permits us to neglect the additional degrees
of freedom occurring in such a case. Thus, from this viewpoint the particle
is still treated as a point-particle. In this paper we intend to show - in
particular - that, unlike the point-charge case,\emph{\ for a finite-size
classical charge of this type the exact RR equation can be explicitly
constructed based on the synchronous Hamilton variational principle. }

\subsection{1a - Motivations and historical background}

The occurrence of self-forces,\ in particular the electromagnetic (EM) one
which is produced by the EM fields generated by the particles themselves, is
an ubiquitous phenomenon which characterizes the dynamics of classical
charged particles. It is well-known that the self-force acts on a (charged)
particle when it is subject also to the action of an arbitrary external
force (Lorentz, 1892 \cite{Lorentz}; see also for example Landau and
Lifschitz, 1957 \cite{LL}). This phenomenon is usually called as \emph{%
radiation reaction} (\emph{RR}) (Pauli \cite{Pauli1958}) or \emph{radiation
damping} \ (see \cite{Feynman1988}), although a distinction between the two
terms is actually made by some authors \cite{Rohrlich2000}.

In classical mechanics the RR problem was first posed by Lorentz in his
historical work (Lorentz, 1985 \cite{Lorentz}; see also Abraham, 1905 \cite%
{Abraham1905}). Traditional approaches are based either on the RR equation
due to Lorentz, Abraham and Dirac (first presented by Dirac in 1938 \cite%
{Dirac1938}), nowadays popularly known as the \emph{LAD equation }or the
equation derived from it by Landau and Lifschitz \cite{LL}, via a suitable
"reduction process", the so-called \emph{LL equation. \ }As recalled
elsewhere (see related discussion in Ref.\cite{Dorigo2008a}) several aspects
of the RR problem - and of the LAD and LL equations - are yet to find a
satisfactory formulation/solution.\ Common feature of all previous
approaches is the adoption of an asymptotic expansion for the EM self-field
(or for the corresponding EM 4-potential), rather than the exact
representation of the force-field. This, in turn, implies that such methods
permit to determine - at most - only an asymptotic approximation for the
(still elusive) exact equation of motion for a charged particle subject to
its own EM self-field.

\subsection{1b - Difficulties with previous RR equations}

Since Lorentz famous paper \cite{Lorentz} many textbooks and research
articles have appeared on the subject of RR. Many of them have criticized
aspects of the RR theory, and in particular the LAD and LL\ equations (for a
review see \cite{Rohrlich1965}, where one can find the discussion of the
related problems). More recently, another equation has been proposed by
Medina \cite{Medina2006}, here denoted as \textit{Medina equation,} which
applies for spherically symmetric and finite-size classical particles. In
these approaches, the charged particles are typically considered
quasi-rigid, i.e., their charge densities are assumed stationary, when seen
with respect to the corresponding particle rest-frame, and eventually also
\emph{point-like}, i.e., both the radii of the mass (if larger than zero)
and of charge distributions are assumed much smaller (i.e., infinitesimal)
with respect to any other classical scale-length characterizing the particle
dynamics.

It is often said that current formulations of the RR problem are
unsatisfactory, because of their possible violation of basic principles of
classical dynamics as well as for some of their properties. These include in
particular:

\begin{itemize}
\item \textit{for the LAD equation}: 1) The violation of Newton's principle
of determinacy (NPD), because the LAD equation requires the specification of
the initial acceleration$,$ besides the initial state; 2) The existence of
so-called runaway solutions, i.e., solutions which blow up in time. In fact,
if a constant external force is applied one can show that the general
solution of the LAD equations diverges exponentially in the future
(blow-up). 3) For the same reason, the LAD equation violates also another
fundamental principle of classical mechanics, the Galilei principle of
inertia (GPI), according to which an isolated particle must have a constant
velocity in any inertial Galilean frame.

\item \textit{for the LL equation}: 1) The use of an iterative approach for
its derivation (from the LAD equation) does not appear justifiable for fully
relativistic particles. In such a case, in fact, the EM self-force cannot
generally be considered a small perturbation of the external EM force. 2)
The LL equation becomes invalid in the case of sudden forces, i.e., forces
which are not smooth functions of time. 3) The neglect of the EM mass: in
the original derivation of the LL equation, given by Landau and Lifschitz
\cite{LL}, the so-called "EM mass" was ignored, which amounts to neglect all
possible EM relativistic corrections to the inertial mass produced by the EM
self-force. In the framework of classical electrodynamics the latter
position appears unfounded (see discussion in Ref. \cite{Dorigo2008a} and
Ref.A). However, in the formulation [of the LL equation] given by Rohrlich
\cite{Rohrlich2001} this effect has been included.

\item \textit{for both equations}: the derivations of both equations (LAD
and LL) are made under the implicit assumption that all the expansions in
powers used near the particle trajectory are valid for the whole range of
values of particle velocity, in particular, arbitrarily close to that of the
light in vacuum. However, it is easy to see that this is not the case.

\item \textit{for the Medina equation}: the use of a perturbative approach,
in particular to evaluate the RR force in the rest frame. This is, however,
a non-relativistic equation. Therefore, the corresponding relativistic
equation is also necessarily asymptotic in character.
\end{itemize}

In our view this clearly indicates that the route to the solution of the RR
problem should be based on the search of the exact relativistic RR equation,
i.e., the construction of a \textit{non-perturbative equation of motion for
a particle in the presence of its EM self-field}.

\subsection{1c - The search of an exact RR equation}

A critical aspect of the RR problem is, however, related to the search of
the \textit{exact relativistic RR equation for classical charged particles,}
in the sense specified above\textit{. }Despite previous attempts, this
equation is still missing. \ \ As far as the LAD equation is concerned \
this is obvious because to obtain it the EM self-field is usually evaluated
by means of an asymptotic expansion. This is true, of course, also for the
LL equation, which according to Rohrlich should be considered as the "exact"
relativistic equation of motion for a classical point-like
spherically-symmetric charge, having a charge distribution with an
infinitesimal radius $\sigma $ \cite{Rohrlich2001} (in this case the
equation is intrinsically asymptotic since it depends on the infinitesimal
parameter $\sigma $).

This feature - as pointed out in Ref.A\ - is also reflected by the
circumstance that these equations are \textit{non-variational }\cite%
{Dorigo2008a}\textit{,} i.e., they do not admit a variational formulation,
at least in the customary sense of the standard Hamilton principle, used in
classical mechanics and electrodynamics \cite{Goldstein}, i.e., for the
conventional 8-dimensional phase-space spanned by the 4-vectors $\left\{
r^{\mu },u_{\mu }=g_{\mu \nu }dr^{\nu }/ds\right\} ,$ $g_{\mu \nu }$
denoting the (Minkowski) metric tensor. This result is clearly in contrast
to the basic principles \textit{both of classical mechanics and
electrodynamics}. In particular, it conflicts with Hamilton's action
principle, which - under such premises (i.e., the validity of LAD and/or LL
equations) - should actually hold true only in the case of inertial motion
(or neglecting altogether the EM self-force)! A consequence which follows is
that the dynamics of point-like charged particles described by these
approximate model equations is not Hamiltonian. However, it is not clear
whether this feature is only an accident, i.e., is only due to the
approximations introduced so far, or is actually an intrinsic feature of the
RR problem.

Another key issue is, however, related to the treatment of the RR problem
for point-particles in a proper sense, and in particular to the conditions
of validity of the relativistic Hamilton variational principle \cite%
{Goldstein} in such a case.\ Actually, difficulties with the treatment of
point-particles in classical electrodynamics and general relativity have
been known for a long time.\ They are due to intrinsic divergences produced
by the EM self-field \cite{Feynman1988-b}. In fact one can show that this
problem is ill-posed since the self-fields diverge in the neighborhood of a
point-particle's world line. \ For this reason in the past several authors,
including Born and Infeld, Dirac, Wheeler and Feynman (see discussion in Ref.%
\cite{Feynman1988}), tried to modify classical electrodynamics in an effort
to eliminate all divergent contributions arising due to EM
self-interactions. This is the so-called regularization problem for
point-particles, based on the introduction of suitable modifications of
Maxwell's electrodynamics.

There is an extensive literature devoted to possible ways to achieve this
goal. These theories either directly introduce 'ad hoc' modified definitions
for the EM self-force (or of the EM self 4-potential) or introduce axiomatic
approaches involving modifications of classical electrodynamics. Examples of
the first type is provided by Dirac \cite{Dirac1938} and Dewitt and Brehme
\cite{DeWitt1960} who determined the RR self-force for a point particle
belonging respectively to the Minkowski and curved space-times by imposing
local energy conservation on a tube surrounding the particle's world line
and subtracting the infinite contributions to the force through a so-called
mass renormalization scheme. More recently, Ori \cite{Ori2004} who suggested
a regularization scheme involving averaging of multipole moments. Another
attempt is based on the adoption of an axiomatic approach in order to
produce the general equation of motion for a point particle coupled to a
scalar field.\ In recent years several different methods have been proposed
for calculating the motion of a point particle coupled to its EM self-fields
(for a review and references on the subject see for example \cite{Quinn2001}%
). Finally, still another possible strategy involves introducing appropriate
modifications of the EM self 4-potential. Typically this is done\ (see for
example Rohlich \cite{Rohlich1964}) by assuming that there exists a
decomposition of the EM field, whereby each particle "feels" only the action
of external particles and of a suitable part of the EM self-field. While
this decomposition becomes clearly questionable for finite-size particles,
its consistency with first principles - and in particular with standard
quantum mechanics - seems dubious, to say the least \cite{Feynman1988}.

Another possible approach for the search of an exact RR equation is
represented by the description of classical charges by means of finite-size
extended particles. An example of this type is provided by Medina \cite%
{Medina2006}, who investigated the dynamics of a point particle
characterized by an arbitrary spherically-symmetric charge. In his approach
a formal integral representation for the RR force in the particle rest frame
is achieved. This result is used to extrapolate the same force for
point-like particles and to evaluate the general form of the RR 4-force in
an arbitrary reference frame, thus yielding an approximate representation of
the relativistic RR equation (\emph{Medina RR equation}). By doing so,
however, an asymptotic approach is inevitably adopted again. Another
interesting feature of the Medina's approach is that the extended-phase
space variational approach is achieved \cite{Medina2007} by introducing an
acceleration-dependent Lagrangian function. \

In this paper, we intend to follow a similar route choosing, however, to
consider: 1) \emph{spherical-shell charges} and 2) \emph{the adoption, from
the beginning, of a relativistic variational approach based on the customary
phase-space Hamilton variational principle}.\ As we intend to prove in the
following, this permits us to obtain \emph{an exact, i.e., non-asymptotic,
RR equation}.\emph{\ }

\subsection{1d - Main results}

\emph{In this paper we intend to pose, for classical finite-size charged
particles represented by shell-charges, the problem of the construction of
the exact RR equation, in the sense indicated above. We want to show that
its explicit construction can be achieved in the framework of classical
electrodynamics, based on a straightforward generalization of Hamilton
variational principle.} The approach is based on the adoption of the
relativistic (hybrid) synchronous Hamilton variational principle recently
pointed out \cite{Cremaschini2006}. Its basic feature is that it can be
expressed virtually in terms of arbitrary "hybrid" variables (i.e.,
generally non-Lagrangian and non-canonical variables). The traditional
approach, valid for point-particles, is extended to finite-size
spherical-shell charges, by taking into account the contribution of the
retarded EM self-potential generated by the particles themselves. Thus,
based on the construction of the Euler-Lagrange equations stemming from the
variational principle, the \emph{exact} relativistic equation of motion for
a charged particle of this type, immersed in a prescribed EM field \emph{and
subject to the simultaneous action of its EM self-field}, can be achieved
explicitly in this way (THM.1-THM.3). In particular, it is found that the%
\emph{\ exact RR equation} in covariant form is (see THM's.1 and 2):

\begin{equation}
m_{o}cdu_{\mu }(s)=\frac{q}{c}\overline{F}_{\mu }^{(ext)\nu }dr_{\nu }(s)+ds%
\overline{G}_{\mu }.  \label{Eq.RR-1}
\end{equation}%
Here $\overline{F}_{\mu }^{(ext)\nu }$ is the surface-average Faraday tensor
- acting on a point particle located at the 4-position $r\equiv \left\{
r^{\mu },\mu =0,3\right\} $ - which is generated by the external EM field.
In particular, the surface-averaging operator acting on a smooth
position-dependent function $A,$ and denoted as $\overline{A},$ is defined
according to Appendix A [see Eq.(\ref{surface average -2})].

Moreover, $\overline{G}_{\mu }$ is the (surface-average) \emph{RR 4-vector }%
produced by the EM self-field and due to the action of the particle on
itself. The rest of the notation is standard. Thus, $c$ is the speed of
light in vacuum, $m_{o}$ and $q$ are respectively the inertial rest-mass and
charge of the particle, $r^{\mu }(s)\equiv r^{\mu }(t(s))$ denotes its
position 4-vector parametrized in terms of the arc lengths $s$ and $u^{\mu
}(s)=\frac{dr^{\mu }(s)}{ds}$ is the corresponding 4-velocity. The explicit
form of $\overline{G}_{\mu }$ is found to be (see THM.2)%
\begin{eqnarray}
&&\overline{G}_{\mu }=2c\left( \frac{q}{c}\right) ^{2}\frac{1}{\left[
R^{\prime \alpha }u_{\alpha }(t)\right] ^{2}}\left[ \frac{dr_{\mu
}(t-t_{ret})}{ds}+\right.  \label{Eq.RR-2} \\
&&\left. -R_{\mu }^{\prime }\frac{u_{k}(t)\frac{dr^{k}(t-t_{ret})}{ds}}{%
R^{\prime \alpha }u_{\alpha }(t)}\right] .  \notag
\end{eqnarray}%
Here $t-t_{ret}$\ is the retarded time, with $t_{ret}$ denoting a suitable
delay-time [see Eq.(\ref{Eq.----21})],\ while $R^{\prime \alpha }=r^{\alpha
}(t)-r^{\alpha }(t-t_{ret})$ and $r^{\alpha }(t-t_{ret})$ is the 4-position
vector evaluated at the retarded time $t^{\prime }.$ As a consequence of Eq.(%
\ref{Eq.RR-2}) the properties of $\overline{G}_{\mu }$ can be immediately
established (see THM.3). \ In particular, $\overline{G}_{\mu }$ depends,
besides $r^{\mu }(s)$ and $u^{\mu }(s)$ evaluated at the local time $t=t(s),$
also on the 4-position and 4-velocity [of the particle itself], \emph{%
evaluated at the retarded time} $t^{\prime },$ i.e., $r^{\mu }(t-t_{ret})$
and $\frac{dr^{\mu }(t-t_{ret})}{ds}.$ It follows that $\overline{G}_{\mu }$
is a smooth function which is generally defined everywhere in a suitable
extended phase space.\ Hence, the RR equation Eq.(\ref{Eq.RR-1}) is a
\textit{retarded second-order ordinary differential equation}. As a main
consequence, the equation, together with the initial conditions
\begin{eqnarray}
r^{\mu }(s_{o}) &=&r_{o}^{\mu },  \label{initial conditions-1} \\
u^{\mu }(s_{o}) &=&u_{o}^{\mu },  \label{initial conditions-2}
\end{eqnarray}%
prescribed so that $u_{o}^{\mu }u_{o\mu }=1$, defines locally a well-posed
problem (THM.1). In addition, its solution results consistent with all basic
principles of classical mechanics, including the principles of Galilei
inertia, Newton determinacy and Einstein causality (THM.3).

To gain deeper insight and to allow comparisons with previous approaches,
various asymptotic approximations and limits are considered in the sequel.
These include: 1) the proof of the non-existence of the point-particle limit
for the present theory (see THM.4), i.e., that the exact RR equation is not
defined in such a case; 2) the "short-time" asymptotic approximation for the
RR equation obtained in the so-called "short-time" ordering (see THM.5).
This is obtained by introducing a Taylor expansion in terms of the
dimensionless ratio $\xi \equiv \left( t-t^{\prime }\right) /t>0$ $($\emph{%
delay-time ratio}$),$ to be assumed infinitesimal; 3) the
weakly-relativistic approximation for the RR equation, obtained by
introducing a Taylor expansion in terms of the dimensionless ratio $\beta
\equiv v(t)/c,$ again to be considered infinitesimal (as appropriate for the
description of non-relativistic particle dynamics; see THM.6).

The analysis is useful to assess the accuracy and limits of validity of the
customary LAD equation, either in the relativistic or weakly-relativistic
descriptions. In both cases it is found that the LAD equation (as well as
the related LL equation) provided, at most, only an asymptotic approximation
to the exact RR equation (\ref{Eq.RR-1}). This conclusion can typically be
reached, however, only provided the external EM field, defined in terms of
the Faraday tensor $F_{\mu }^{(ext)\nu }$, is a suitably smooth function of
the particle proper time $\tau \equiv s/c.$ In particular, both LAD and LL
equations may not be valid for "sudden forces", i.e., external fields which
are locally discontinuous with respect to $\tau $.

\subsection{1e - Scheme of presentation}

The scheme of the presentation is as follows. In Sections 1 and 2 a brief
overview of previous treatments is given in order to analyze the intrinsic
difficulties met by previous point-charges descriptions for the RR problem.
In the subsequent sections (Sec. 3 to 7) the new treatment which applies for
finite-size charges is presented. In particular:

\begin{itemize}
\item In Sec.3 the exact EM 4-potential generated by a finite-size spherical
shell is evaluated.

\item In Sec.4 the Hamilton synchronous variational principle for a
finite-size charge is developed and an explicit form of the RR equation is
obtained (see THM.1). In particular, it is proven that the resulting
relativistic RR equation is a second-order ordinary differential equation
which defines a well-posed problem, i.e., that the solution of the
corresponding initial-value problem locally exists and is unique.

\item As a consequence (Sec.5), the 4-vector\ ($\overline{G}_{\mu }$) is
introduced which describes the generalized Lorentz force acting on the
particle generated by its EM self-field (see THM.2).

\item In Sec.6 the main properties of $\overline{G}_{\mu }$ are
investigated. As a result it is proven that the RR equation is consistent
with all basic principles of classical mechanics (THM.3).

\item In Sec.7 the non-existence of the point-particle limit [for the RR
equation] is proven (THM.4).
\end{itemize}

In Sec.8 the short-time approximation for $\overline{G}_{\mu }$ is obtained.
The resulting asymptotic RR equation is found consistent with the customary
relativistic LAD equation (THM.5). Finally, in Sec.9 possible
weakly-relativistic approximations of the RR equation are discussed. Also in
this case, the resulting RR equation can be realized, unlike the customary
weakly-relativistic LAD equation, by means of a second-order differential
equation (THM.6).

\section{2 - The impossibility in classical electrodynamics of a variational
description for classical point-charges}

A corner-stone of classical mechanics is represented by the Hamilton
variational principle, which permits to determine the coupled set of
equations formed by the particle dynamical equations and Maxwell's equations
\cite{LL,Goldstein}. As a consequence, both the particle state and the EM
field in which the particle is immersed are uniquely determined by means of
this variational principle. The choice of the dynamical variables which
define the particle state remains in principle arbitrary. Thus, they can
always be represented by so-called "hybrid" variables, i.e., superabundant
variables which generally do not define a Lagrangian state. This implies,
thanks to Darboux theorem, that it should always be possible to identify
them locally with canonical variables. As a basic consequence, classical
systems of charged particles are expected to define Hamiltonian systems,
i.e., their canonical states should be extrema of the corresponding
Hamiltonian action, while the corresponding particle dynamics, provided by
the Euler-Lagrange equations determined by the same variational principle,
necessarily should coincide with Hamilton's equations of motion.

Nevertheless, it is easy to prove that for charged point-particles the
Hamilton principle fails (see Ref.A). In fact, one can show that, if the
Hamilton principle is expressed via a synchronous hybrid variational
principle \cite{Cremaschini2006}, the point-charge action integral can be
written in the form (here the notation is given according to Ref.A)
\begin{eqnarray}
&&S_{1}(r^{\mu },u_{\mu },\chi )=  \label{Eq.---1b-} \\
&&\left. =\int_{1}^{2}\left( m_{o}cu_{\mu }+\frac{q}{c}A_{\mu }(r)\right)
dr^{\mu }+\right.   \notag \\
&&+\int_{s_{1}}^{s_{2}}ds\chi (s)\left[ u_{\mu }(s)u^{\mu }(s)-1\right]
\notag
\end{eqnarray}%
(Hamiltonian action), which is applicable if the 4-potential $A_{\mu }$ is
considered prescribed. It is immediate \ to prove that the functional is
actually \emph{not-defined.} The reason is due to the intrinsic divergences
appearing in the point-particle self 4-potential $A_{\mu }^{(self)}$ (see
Appendix B)$.$ In fact, due to the superposition principle the EM
4-potential $A_{\mu }(r)$ can always be represented in terms of the
fundamental decomposition
\begin{equation}
A_{\mu }=A_{\mu }^{(self)}+A_{\mu }^{(ext)},  \label{Eq.----2}
\end{equation}%
where $A_{\mu }^{(self)}$\ and \ $A_{\mu }^{(ext)}$ denote respectively the
point-particle self 4-potential and the external 4-potential. In
particular,\ by assumption $A_{\mu }^{(self)}\equiv A_{\mu }^{(self)}(r(s))$
is a solution of the Maxwell's equations which in flat space-time are given
by%
\begin{equation}
\left. \partial _{\nu }F^{\nu \mu (self)}=\frac{4\pi }{c}j^{\mu },\right.
\label{Eq.----3}
\end{equation}%
with $\ j^{\mu }(r^{\nu })=\int_{s_{1}}^{s_{2}}ds^{\prime }u^{\mu
}(s^{\prime })\delta ^{(4)}\left( r-r(s^{\prime })\right) $ denoting the
4-current carried by the point charge. Hence, in the functional $%
S_{1}(r^{\mu },u_{\mu },\chi ),$ the 4-vector function $A_{\mu }$ must be
considered a prescribed function of the varied 4-vector $r^{\mu }(s)$. \
Invoking the causality principle, the explicit form of $A_{\mu }^{(self)}$
for a point particle immersed in the Minkowski space-time $M^{4}\equiv
%TCIMACRO{\U{211d} }%
%BeginExpansion
\mathbb{R}
%EndExpansion
^{4}$ can be easily recovered (see Appendix B) and is provided by the
well-known retarded EM 4-potential (in covariant form)%
\begin{equation}
A_{\mu }^{(self)}(r)=\frac{q}{c}\frac{u_{\mu }(t^{\prime })}{R^{\alpha
}u_{\alpha }(t^{\prime })},  \label{Eq.----4}
\end{equation}%
which can be represented in the equivalent integral form given by Eq.(\ref%
{10}). Here $R^{\mu },$ $u_{\mu }(t^{\prime }),u^{\mu }(t^{\prime })$ and $%
t^{\prime }$ are respectively the bi-vector $R^{\mu }\equiv r^{\mu
}-r^{\prime \mu },$ with $r^{\prime \mu }\equiv r^{\mu }(t^{\prime }),$ the $%
4-$velocity $u^{\mu }(t^{\prime })\equiv \frac{d}{ds^{\prime }}r^{\mu
}(t^{\prime })=\gamma ^{\prime }\frac{d}{dt^{\prime }}r^{\mu }(t^{\prime })$
and its covariant components $u_{\mu }(t^{\prime })$, while $t^{\prime }$ is
a suitable retarded time. In particular this is defined so that
\begin{equation}
\left. t-t^{\prime }=\frac{\left\vert \mathbf{r}-\mathbf{r}^{\prime
}\right\vert }{c}\right. ,  \label{Eq.----5}
\end{equation}%
where $\mathbf{r}^{\prime }\equiv \mathbf{r}(t^{\prime }).$ We now notice
that for an arbitrary varied curve $r(s),$ the inf of $t-t^{\prime }$ is
generally not strictly positive in the case of a point-charge. As a
consequence, the contributions carried by $A_{\mu }^{(self)}$ in the
functional $S_{1}(r^{\mu },u_{\mu },\chi )$\ contain essential divergences.
This means that, when the self 4-potential $A_{\mu }^{(self)}$ is properly
taken into account in the point-charge action functional $S_{1}(r^{\mu
},u_{\mu },\chi ),$ the functional cannot actually be defined.

\section{3 - The retarded EM self 4-potential of a finite-size charge}

A prerequisite for the subsequent developments is the determination of the
EM self-potential ($A_{\mu }^{(self)}$) produced by a prescribed charge
distribution. As indicated above, in this paper we wish to consider the case
of a classical particle characterized by point-particle mass and -
respectively - finite-size charge distributions. For definiteness, here we
shall determine the EM 4-potential generated by a finite-size
spherical-shell particle immersed in the Minkowski space-time$.$ In
particular, we assume that when observed with respect to the particle
rest-frame the charge density takes the form
\begin{equation}
\rho (\mathbf{r},t)=\frac{q}{4\pi \sigma ^{2}}\delta (\left\vert \mathbf{r}-%
\mathbf{r}(t)\right\vert -\sigma ).  \label{charge  density}
\end{equation}%
First, let us evaluate the retarded electrostatic (ES) potential generated
by $\rho (\mathbf{r},t)$ and measured at a position $\mathbf{r}$ defined in
such a frame. This is manifestly defined as
\begin{equation}
\Phi ^{(self)}(\mathbf{r},t)=\int d^{3}r^{\prime }\frac{1}{R}\rho (\mathbf{r}%
,t-\frac{R}{c}),  \label{A-0}
\end{equation}%
with $R\equiv \left\vert \mathbf{R}\right\vert $ and $\mathbf{R}=\mathbf{r-r}%
^{\prime }.$ It is well known that $\Phi ^{(self)}(\mathbf{r},t)$ can be
determined conveniently by introducing an expansion in Legendre polynomials
for the integrand $\frac{1}{R}\rho (\mathbf{r},t-\frac{R}{c})$. As a result
one can readily show that \emph{for a finite-size spherical-shell charge the
retarded ES potential is }(see \ for example\emph{\ }\cite{Michelitsch2005})%
\begin{equation}
\Phi ^{(self)}(\mathbf{r},t)=\left\{
\begin{array}{ccc}
\frac{q}{R} &  & R\geq \sigma \\
\frac{q}{\sigma } &  & R<\sigma ,%
\end{array}%
\right.  \label{A-2'}
\end{equation}%
\emph{\ }where\emph{\ }%
\begin{eqnarray}
R &\equiv &\left\vert \mathbf{R}\right\vert ,  \label{A-4} \\
\mathbf{R} &=&\mathbf{r}-\mathbf{r}(t-\frac{\left\vert \mathbf{r}-\mathbf{r}%
(t-\frac{R}{c})\right\vert }{c}).  \label{A-5}
\end{eqnarray}%
Therefore, in the internal domain ($R<\sigma $) the EM self-potential does
not produce any self-field. Instead, in the external domain\ ($R\geq \sigma $%
) its expression is the same as that produced by a point-charge. In both
cases the ES potential is manifestly spherically symmetric, therefore it
follows by construction that in the rest frame:%
\begin{equation}
\overline{\Phi }^{(self)}(\mathbf{r,}t)=\Phi ^{(self)}(\mathbf{r},t),
\end{equation}%
where $\overline{\Phi }^{(self)}(\mathbf{r,}t)$ is the surface-average (\ref%
{surface-average}). \ The corresponding expression of the EM 4-potential in
a moving frame can be easily obtained by applying a Lorentz transformation.
In particular, since in this case the external domain is defined by the
inequality $R^{\alpha }R_{\alpha }\geq \sigma ^{2}$ the corresponding
surface-average EM self 4-potential $\overline{A}_{\mu }^{(self)}$\ is given
again by Eq.(\ref{Eq.----4}), namely

\begin{equation}
\overline{A}_{\mu }^{(self)}(r)=\frac{q}{c}\frac{u_{\mu }(t^{\prime })}{%
R^{\alpha }u_{\alpha }(t^{\prime })}.  \label{differential representation}
\end{equation}%
Instead, in the internal domain ($R^{\alpha }R_{\alpha }<\sigma ^{2}$) there
results necessarily $\overline{A}_{\mu }^{(self)}=const.,$ so that $%
\overline{F}^{\mu \nu (self)}\equiv 0$ in this subset. As a further
consequence, if $r^{\alpha }$ is the 4-position vector of the
point-particle, there results (see Appendix C)
\begin{equation}
\overline{A}_{\mu }^{(self)}(r)=\frac{2q}{c}\int_{_{1}}^{2}dr_{\mu }^{\prime
}\delta (R^{\alpha }R_{\alpha }-\sigma ^{2}),
\label{integral representation}
\end{equation}%
where $R^{\alpha }=r^{\alpha }-r^{\prime \alpha }.$

\section{4 - The RR equation for a finite-size charge}

In this section we wish address the key issue posed in this paper, i.e., the
problem of the explicit construction of the relativistic RR equation for a
finite-size spherical shell charge. Here we intend to prove that, as earlier
pointed out in Ref.A, this goal can be uniquely established based on a
suitable formulation of the Hamilton variational principle. More precisely,
we intend to prove that:

\begin{itemize}
\item the exact RR equation can be obtained by making use of a suitably
modified form of the synchronous Hamilton variational principle appropriate
for finite-size charges (see THM.1);

\item the solution of the related initial-value problem exists and is
unique, i.e., the RR equation defines a well-posed problem (THM.1).
\end{itemize}

\subsection{4a - Treatment of finite-size particles}

First, let us generalize the \emph{Hamilton action functional [Eq.(\ref%
{Eq.---1b-})] to treat finite-size particles. } This is obtained formally by
introducing in $S_{1}(r^{\mu },u_{\mu },\chi )$ the replacements \
\begin{eqnarray}
ds &\rightarrow &W(r,s)\frac{d\Omega }{\sqrt{-g}}, \\
dr^{\mu } &\rightarrow &\frac{dr^{\mu }}{ds}W(r,s)\frac{d\Omega }{\sqrt{-g}},
\end{eqnarray}%
where $W(r,s)$ is the so-called "wire function", generally to be identified
with a suitable distribution. For a generic $W(r,s)$ the appropriate form of
the variational functional (to be expressed again in synchronous form \cite%
{Cremaschini2006}) becomes%
\begin{eqnarray}
&&\left. S_{1}(r^{\mu },u_{\mu },\chi )=\right.  \label{Eq.---10} \\
&&\left. =\int \frac{d\Omega }{\sqrt{-g}}W(r,s)\left( m_{o}cu_{\mu }+\frac{q%
}{c}A_{\mu }(r)\right) \frac{dr^{\mu }}{ds}+\right.  \notag \\
&&+\int_{s_{1}}^{s_{2}}\frac{d\Omega }{\sqrt{-g}}W(r,s)\chi (s)\left[ u_{\mu
}u^{\mu }-1\right] .  \notag
\end{eqnarray}

\subsection{4b - The wire function of a spherical-shell charge}

Let us now consider, in particular, the case of a spherical-shell
charge, while requiring that the mass is still point-wise
localized, i.e., it is a point-particle (see discussion in Sec.1a)
with 4-position $r^{\mu }(s)$ and 4-velocity $u^{\mu }(s)$. \ To
obtain the appropriate representation of the wire function in this
case, let us introduce the coordinate transformation $r^{\mu
}\rightarrow \left( s,\xi ^{1},\xi ^{2},\rho \right) .$ Here $s$
is the arc length along the particle world line, $\xi ^{1}$ and
$\xi ^{2}$ are two curvilinear angle-like coordinates on the
surface $\partial \Omega _{\sigma }$ and $\rho $ is the 4-scalar
defined so that $\rho ^{2}=\widehat{R}^{\alpha
}\widehat{R}_{\alpha },$ with $\widehat{R}^{\alpha }=r^{\alpha
}-r^{\alpha }(s)$ and $r^{\alpha }(s)$ denoting the 4-position of
the point particle. It follows that the
wire-function for a spherical-shell charge can be defined as%
\begin{equation}
W=\frac{1}{4\pi \sigma ^{2}}\delta \left( \rho -\sigma \right) ,
\label{wire function (general)}
\end{equation}%
while the invariant volume element is $\frac{d\Omega }{\sqrt{-g}}=ds\rho
^{2}d\rho \frac{d\Sigma (n)}{\sqrt{-g}},$ with $\sqrt{-g}=1$ for flat
space-time and $\frac{d\Sigma (n)}{\sqrt{-g}}$ denoting a suitable invariant
surface element. It follows that $W$ is non-zero only if
\begin{equation}
r^{\alpha }=r^{\alpha }(s)+\sigma n^{\alpha }(\xi ^{1},\xi ^{2}),
\label{4-position}
\end{equation}%
where $n^{\alpha }$ is a unit 4-vector ($n^{\alpha }n_{\alpha }=1$)
depending only on $(\xi ^{1},\xi ^{2})$. Thus, in particular, in the
rest-frame of the same particle $W$ takes the form
\begin{equation}
W=\frac{1}{4\pi \sigma ^{2}}\delta \left( \left\vert \mathbf{r}-\mathbf{r}%
(s)\right\vert -\sigma \right) ,  \label{wire-function}
\end{equation}%
while $d\Sigma (n)$ can be identified with the solid angle (surface element
of 3-sphere of unit radius centered at the particle position $\mathbf{r}$), $%
\rho =\left\vert \mathbf{r}-\mathbf{r}(s)\right\vert $ and the 4-vector $n$
reads $n=(0,\mathbf{n}),$ $\mathbf{n}$ denoting the normal unit 3-vector to
the surface $\partial \Omega _{\sigma }$. It follows that for an arbitrary
4-tensor $A(r(s)+\sigma n)$ evaluated at the 4-position (\ref{4-position})
one can define an appropriate surface-average \ (see Appendix A).

\subsection{4c - Spherical-shell charge Hamilton principle}

The appropriate form of the Hamilton action functional for a spherical-shell
charge is found to be given by the following Lemma.

\textbf{LEMMA 1 - Spherical-shell charge action integral}

\emph{For a finite-size spherical-shell charge the Hamiltonian action
integral defined by Eq.(\ref{Eq.---10}) reads:}
\begin{eqnarray}
&&\left. S_{1}(r^{\mu },u_{\mu },\chi )=\right.  \label{Eq.----A.1} \\
&&\left. =\frac{1}{4\pi }\int d\Sigma (n)\int_{1}^{2}\left[ m_{o}cu_{\mu
}(s)\right. +\right.  \notag \\
&&\left. +\left. \frac{q}{c}A_{\mu }^{(ext)}(r(s)+\sigma n)\right] dr^{\mu
}+\right.  \notag \\
&&\left. +\Delta S_{1}(r^{\mu })+\right.  \notag \\
&&\left. +\frac{1}{4\pi }\int d\Sigma (n)\int_{s_{1}}^{s_{2}}\chi (s)\left[
u_{\mu }(s)u^{\mu }(s)-1\right] ds\right.  \notag
\end{eqnarray}

\emph{(Hamilton action integral), where }$\Delta S_{1}(r^{\mu })$ \emph{is}
\emph{the functional carrying the contribution of the EM self 4-potential }
\emph{\ }%
\begin{equation}
\Delta S_{1}(r^{\mu })\equiv \frac{1}{4\pi }\int d\Sigma (n)\int_{1}^{2}%
\frac{q}{c}A_{\mu }^{(self)}(r(s)+\sigma n))dr^{\mu }.
\end{equation}%
\emph{\ In view of Eq.(\ref{integral representation}) and the surface
average (\ref{surface average -2}) there results }%
\begin{eqnarray}
&&\left. \Delta S_{1}(r^{\mu })=\right.  \label{Eq.-----A.2} \\
&&\left. =2\left( \frac{q}{c}\right) ^{2}\int_{_{1}}^{_{2}}dr^{\mu
}\int_{_{1}}^{2}dr_{\mu }^{\prime }\delta (R^{\alpha }R_{\alpha }-\sigma
^{2})\right. .  \notag
\end{eqnarray}%
\emph{\ }

\emph{Proof} - The proof of Eq.(\ref{Eq.----A.1}) follows from the
wire-function functional [Eq.(\ref{wire function (general)})] upon invoking
Eq.(\ref{wire-function}) for the wire function. Instead, the specific form
of the functional $\Delta S_{1}(r^{\mu })$ [Eq.(\ref{Eq.-----A.2})], which
carries the EM self 4-potential, follows invoking the integral
representation (\ref{integral representation}). \textbf{Q.E.D. }

As a basic consequence,\ invoking in particular the\emph{\ }surface-average
of\emph{\ }$A_{\mu }^{(ext)}(r(s)+\sigma n)$ given by\emph{\ }Eq. (\ref%
{surface average -2}), all terms in the integrand of the action functional (%
\ref{Eq.----A.1}) become independent of the surface element $d\Sigma (n).$
Hence, the action functional reduces simply to:
\begin{eqnarray}
&&S_{1}(r^{\mu },u_{\mu },\chi )=  \label{Functional-finale-1} \\
&&\left. =\int_{1}^{2}\left( m_{o}cu_{\mu }(s)+\frac{q}{c}\overline{A}_{\mu
}^{(ext)}(r(s))\right) dr^{\mu }+\right.  \notag \\
&&+2\left( \frac{q}{c}\right) ^{2}\int_{_{1}}^{_{2}}dr^{\mu
}\int_{_{1}}^{2}dr_{\mu }^{\prime }\delta (R^{\alpha }R_{\alpha }-\sigma
^{2})+  \notag \\
&&+\int_{s_{1}}^{s_{2}}\chi (s)\left[ u_{\mu }(s)u^{\mu }(s)-1\right] ds.
\notag
\end{eqnarray}

Let us now prove that the relativistic dynamics of a (finite-size) spherical
shell particle is uniquely prescribed by the Hamilton variational principle
defined in terms of $S_{1}(r^{\mu },u_{\mu },\chi ),$ specified according to
Eq.(\ref{Eq.----A.1}). In particular, in this case, due to the assumption
that the mass of the particle is point-wise localized, the extremal curve
must be necessarily of the form $r\equiv r(s)$ [see Assumption 3 in THM.1].
The following result then holds:

\textbf{THM.1 - Hamilton principle for a spherical-shell charge }

\emph{Let us assume that: 1) the real varied functions }$f(s)\equiv \left[
r^{\mu }(s),u_{\mu }(s),\chi (s)\right] $\emph{\ belong to a suitable
functional class }$\left\{ f\right\} $ \emph{in which end points and
boundaries are kept fixed; 2) the Hamilton action integral }$S_{1}(r^{\mu
},u_{\mu },\chi )$ \emph{defined by Eq.(\ref{Eq.----A.1}) is assumed to
exist for all }$f(s)\in \left\{ f\right\} .$ \emph{Here, }$u^{\mu
}(s)=g^{\mu \nu }u_{\nu }(s),$\emph{\ while }$g^{\mu \nu }=g^{\mu \nu
}(r(s)) $\emph{\ denotes the counter-variant components of the metric
tensor, each one to be considered dependent on the generic varied curve }$%
r(s);$\emph{\ furthermore, }$m_{o}$\emph{\ and }$q$\emph{\ are respectively
the constant rest mass and electric charge of a point particle and }$ds$%
\emph{\ the line element; 3) an extremal curve }$f\in \left\{ f\right\} $
\emph{of }$S_{1}$ \emph{is assumed of the form }$f(s),$ \emph{i.e., to be
independent of }$n$; \emph{4) if }$r(s)$ \emph{is an extremal curve of }$%
S_{1}$ \emph{the line element }$ds$\emph{\ satisfies the constraint }$%
ds^{2}=g_{\mu \nu }(r(s))dr^{\mu }(s)dr^{\nu }(s).$\emph{\ }

\emph{Then it follows that:}

\emph{T1}$_{1})$\emph{\ if the synchronous variations }$\delta f(s)$ \emph{%
[see also Appendix D] are considered as independent, the Euler-Lagrange
equations following from the synchronous variational principle}%
\begin{equation}
\delta S_{1}(r^{\mu },u_{\mu },\chi )=0  \label{Eq.----20A}
\end{equation}%
\emph{\ \ yield identically the RR equation of motion for a finite-size
spherical-shell charged particle, which reads:}
\begin{eqnarray}
&&m_{o}cdu_{\mu }(s)=\frac{q}{c}\overline{F}_{\mu }^{(ext)\nu }(r(s))dr_{\nu
}(s)+  \label{Eq.---------------20B} \\
&&+dr^{k}H_{\mu k},  \notag
\end{eqnarray}%
\emph{where }$\overline{F}_{\mu }^{(ext)\nu }(r(s))$ \emph{is the
surface-average of the Faraday 4-tensor }$\overline{F}_{\mu \nu
}^{(ext)}\equiv \partial _{\mu }\overline{A}_{\nu }^{(ext)}-\partial _{\nu }%
\overline{A}_{\mu }^{(ext)}$\emph{\ evaluated at the 4-position }$r(s)$\emph{%
.} \emph{In addition}, $r^{\mu }\equiv r^{\mu }(t),r^{\prime \mu }\equiv
r^{\mu }(t^{\prime }),$ \emph{while} $u^{\mu }=\frac{dr^{\mu }}{ds}$\emph{\
is the 4-velocity, }$v^{\mu }(t)$ \emph{denotes }$v^{\mu }(t)=\frac{dr^{\mu }%
}{dt}$ \emph{and }$H_{\mu k}$\emph{\ is the function}%
\begin{eqnarray}
&&\left. H_{\mu k}=2\left( \frac{q}{c}\right) ^{2}\left[ \frac{1}{%
c\left\vert (t-t^{\prime })-\frac{1}{c^{2}}\frac{d\mathbf{r}(t^{\prime })}{%
dt^{\prime }}\cdot (\mathbf{r-r}^{\prime })\right\vert }\right. \right. \\
&&\left. \left. \frac{d}{dt^{\prime }}\left\{ \frac{v_{\mu }(t^{\prime
})R_{k}-v_{k}(t^{\prime })R_{\mu }}{c^{2}\left\vert (t-t^{\prime })-\frac{1}{%
c^{2}}\frac{d\mathbf{r}(t^{\prime })}{dt^{\prime }}\cdot (\mathbf{r-r}%
^{\prime })\right\vert }\right\} \right] _{t^{\prime }=t-t_{ret}}=0.\right.
\notag
\end{eqnarray}%
\emph{Finally,\ }$\mathbf{r}\equiv \mathbf{r}(t)$ \emph{and} $\mathbf{r}%
^{\prime }\equiv \mathbf{r}(t^{\prime }),$ \emph{while }$t^{\prime
}=t-t_{ret}$\emph{\ denotes the retarded time and }$t_{ret}$\ \emph{a
suitable delay-time; }

\emph{T1}$_{2})$\emph{\ the delay-time }$t_{ret}$\emph{\ is the positive
root of the equation }%
\begin{equation}
R^{\alpha }R_{\alpha }=\sigma ^{2}  \label{Eq.-----21A}
\end{equation}%
\emph{(delay-time equation)} \emph{which is}
\begin{eqnarray}
&&\left. t_{ret}(t)\equiv t-t^{\prime }=\right.  \label{Eq.----21} \\
&=&\frac{1}{c}\sqrt{\left[ \mathbf{r}(t)\mathbf{-r}(t-t_{ret}(t))\right]
^{2}+\sigma ^{2}}>0;  \notag
\end{eqnarray}

\emph{T1}$_{3})$\emph{\ let us require that the 4-vector-field }$\overline{A}%
_{\mu }^{(ext)}(r)$ \emph{is suitably smooth in the whole Minkowski
space-time }$M^{4}$\emph{, i.e., is at least }$C^{(2)}(M^{4});$\emph{\ then
the initial-value problem set by the Euler-Lagrange equation (\ref%
{Eq.---------------20B}), with the initial conditions }%
\begin{equation}
\mathbf{x}(t_{o})=\mathbf{x}_{o},  \label{Eq.-----------21C}
\end{equation}%
\emph{[where }$x(t_{o})\equiv \left[ r^{\mu }(t_{o}),u_{\mu }(t_{o})\right] $%
\emph{\ and }$x_{o}\equiv \left[ r_{o}^{\mu },u_{\mu o}\right] $\emph{\
denotes a suitable initial state], is locally well-posed. }

\emph{Proof} - \emph{T1}$_{1})$ It is immediate to construct explicitly the
Euler-Lagrange equations of the Hamilton action $S_{1}(r^{\mu },u_{\mu
},\chi )$. In fact, first, since $\frac{\partial }{\partial u^{\mu }}\delta
(R^{\alpha }R_{\alpha }-\sigma ^{2})=\frac{\partial }{\partial u^{\prime \mu
}}\delta (R^{\alpha }R_{\alpha }-\sigma ^{2})\equiv 0$, the variations with
respect to $\chi (s)$ and $u_{\mu }$ deliver respectively%
\begin{eqnarray}
&&\left. u_{\mu }(s)u^{\mu }(s)-1=0,\right. \\
&&\left. m_{o}cdr^{\mu }+2\chi (s)u^{\mu }(s)ds=0,\right.
\label{Eq.-------------------32A}
\end{eqnarray}%
while it must result for consistency $2\chi (s)=-m_{o}c$ (as in the case in
which $A_{\mu }^{(self)}$ is assumed to vanish identically). To reach Eq.(%
\ref{Eq.---------------20B}), instead, let us invoke Lemma 2 [see Appendix
D]. Then, thanks to assumption 3), the variation with respect to $r^{\mu }$
can easily be proven to yield the Euler-Lagrange equation defined by Eq.(\ref%
{Eq.---------------20B}). Together with Eq.(\ref{Eq.-------------------32A}%
), this manifestly defines the \emph{RR equation}, i.e., \emph{the} \emph{%
exact relativistic equation of motion for a point charge subject to the
simultaneous action of a prescribed external EM field and of its self-EM
field. }

\emph{T1}$_{2})$ Recalling that in the Minkowski metric the retarded-time
equation [Eq.(\ref{Eq.-----21A})] reads
\begin{equation}
R^{\alpha }R_{\alpha }=c^{2}(t^{\prime }-t)^{2}-(\mathbf{r-r}^{\prime
})^{2}=\sigma ^{2},  \label{Eq.-----24}
\end{equation}%
with $R^{\alpha }=r^{\alpha }(t)-r^{\alpha }(t^{\prime })$ and $\mathbf{r}=%
\mathbf{r}(t),\ \mathbf{r}^{\prime }=\mathbf{r}(t^{\prime }),$ the proof of
Eq.(\ref{Eq.----21}) is straightforward$.$

\emph{T1}$_{3})$ Finally, it is immediate to show that the problem defined
by Eq.(\ref{Eq.---------------20B}), together with the initial conditions
defined by Eq.(\ref{Eq.-----------21C}), admits a local existence and
uniqueness theorem (fundamental theorem). In fact it is obvious that Eq.(\ref%
{Eq.---------------20B}) can be cast in the form of a delay-differential
equation, i.e.,%
\begin{equation}
\frac{d\mathbf{x}(t)}{dt}=\mathbf{X(x}(t),\mathbf{x}(t-t_{ret}),t),
\label{Eq.------------------24}
\end{equation}%
where $\mathbf{x}(t)$ and $\mathbf{x}(t-t_{ret})$ denote respectively the
"instantaneous" and "retarded" states $\mathbf{x}(t)\equiv \left[ r^{\mu
}(t),u_{\mu }(t)\right] $ and $\mathbf{x}(t-t_{ret})\equiv \left[ r^{\mu
}(t-t_{ret}),u_{\mu }(t-t_{ret})\right] ,$ while $\mathbf{X(x}(t),\mathbf{x}%
(t-t_{ret}),t)$ is a suitable $C^{(2)}$ real vector field depending smoothly
on both of them. It is manifest that the fundamental theorem holds for Eqs.(%
\ref{Eq.-----------21C})-(\ref{Eq.------------------24}). In fact, by
considering [in $\mathbf{X}$] $\mathbf{x}(t-t_{ret})$ as a prescribed
function of time, the previous equation recovers the canonical form
\begin{equation}
\frac{d\mathbf{x}(t)}{dt}=\widehat{\mathbf{X}}\mathbf{(x}(t),t),
\end{equation}%
with $\widehat{\mathbf{X}}\mathbf{(x}(t),t)$ denoting the corresponding $%
C^{(2)}$ real vector field. This proves the statement. \textbf{Q.E.D.}

\section{5 - Determination of the RR 4-vector\emph{\ }$\overline{G}_{\protect%
\mu }$}

A basic consequence of THM.1 is that the RR equation can be expressed in
covariant form. This permits us to identify the \emph{RR 4-vector }$%
\overline{G}_{\mu },$ which represents \emph{the (generalized) Lorentz force
produced on a charged particle by its EM self-field}. Here we intend to
show, in particular, that $\overline{G}_{\mu }$ can be expressed in
covariant form and uniquely parametrized in terms of the proper length $s,$
defined at the point-particle 4-position vector $r^{\mu }$. The main result
is represented by the following theorem which provides also an explicit
representation of the 4-vector $\overline{G}_{\mu }$.\bigskip

\textbf{THM.2 - Covariant representation of }$\overline{G}_{\mu }$

\emph{For the Minkowski metric the covariant RR equation reads }%
\begin{eqnarray}
&&m_{o}cdu_{\mu }(s)=\frac{q}{c}\overline{F}_{\mu }^{(ext)\nu }(r(s))dr_{\nu
}(s)+  \label{Eq.----29} \\
&&\left. +\overline{G}_{\mu }ds.\right.  \notag
\end{eqnarray}%
\emph{Here the 4-vector }$\overline{G}_{\mu }\equiv \left( G_{o},\mathbf{G}%
\right) $\emph{\ is defined as:}

\begin{eqnarray}
&&\overline{G}_{\mu }=2c\left( \frac{q}{c}\right) ^{2}u^{k}(s)\left[ \frac{1%
}{R^{\alpha }u_{\alpha }(t^{\prime })}\right.  \label{Eq.-----30} \\
&&\left. \frac{d}{ds^{\prime }}\left\{ \frac{\frac{d}{ds^{\prime }}r_{\mu
}(t^{\prime })R_{k}-\frac{d}{ds^{\prime }}r_{k}(t^{\prime })R_{\mu }}{%
R^{\alpha }u_{\alpha }(t^{\prime })}\right\} \right] _{t^{\prime }=t=t_{ret}}
\notag
\end{eqnarray}%
\emph{(covariant representation with respect to }$s^{\prime }$\emph{). Here }%
$s$\emph{\ and }$s^{\prime }$\emph{\ are defined respectively by }$ds=cdt%
\sqrt{1-\beta ^{2}(t)}$\emph{\ and }$ds^{\prime }=cdt^{\prime }\sqrt{1-\beta
^{2}(t^{\prime })},$\emph{\ where }$t^{\prime }=t-t_{ret}$\emph{\ is the
retarded time and }$\beta ^{2}(t)=\frac{1}{c^{2}}\left( \frac{d\mathbf{r}(t)%
}{dt}\right) ^{2}$. \emph{An equivalent representation of \ }$\overline{G}%
_{\mu }$\emph{\ in terms of the particle arc length }$s$\emph{\ is:}%
\begin{eqnarray}
&&\overline{G}_{\mu }=  \label{Eq.--------30bis} \\
&&\left. =2c\left( \frac{q}{c}\right) ^{2}\frac{1}{\left[ R^{\prime \alpha
}u_{\alpha }(t)\right] ^{2}}\left[ \frac{dr_{\mu }(t-t_{ret})}{ds}+\right.
\right.  \notag \\
&&\left. -R_{\mu }^{\prime }\frac{u_{k}(t)\frac{dr^{k}(t-t_{ret})}{ds}}{%
R^{\prime \alpha }u_{\alpha }(t)}\right]  \notag
\end{eqnarray}%
\emph{(covariant representation with respect to }$s$\emph{), where }$%
R^{\prime \alpha }=r^{\alpha }(t)-r^{\alpha }(t-t_{ret}).$\emph{\ This can
be proven to yield also a parametric representation of }$\overline{G}_{\mu }$%
\emph{\ in terms of }$s.$

\emph{Proof} - The proof of the first covariant representation of $\overline{%
G}_{\mu }$\ [given by Eq.(\ref{Eq.-----30})] follows immediately. In fact,
by definition there results $\frac{d}{dt^{\prime }}=\frac{c}{R^{\prime
\alpha }u_{\alpha }(t^{\prime })}\frac{d}{ds^{\prime }},$ where $u_{\alpha
}(t^{\prime })=\gamma (t^{\prime })v_{\alpha }(t^{\prime }),$ with $\gamma
(t^{\prime })=1/\sqrt{1-\beta ^{2}(t^{\prime })}$ and $v_{\alpha }(t^{\prime
})$ denoting $v_{\alpha }(t^{\prime })=\frac{dr^{\alpha }(t^{\prime })}{%
dt^{\prime }}.$ Instead, to prove the representation (\ref{Eq.--------30bis}%
) we first notice that by construction $d(R^{\alpha }R_{\alpha })=0.$ Hence
the two differential constraints $dr^{k}(t)R_{k}=dr^{k}(t^{\prime })R_{k}$
and $\frac{1}{R^{\alpha }u_{\alpha }(t^{\prime })}\frac{d}{ds^{\prime }}=%
\frac{1}{R^{\alpha }u_{\alpha }(t)}\frac{d}{ds}$ (see also Lemma 3 in
Appendix D) must be fulfilled too. This implies that the following
differential identity must hold
\begin{eqnarray}
&&\left. \frac{\frac{d}{ds^{\prime }}r_{\mu }(t^{\prime })R_{k}-\frac{d}{%
ds^{\prime }}r_{k}(t^{\prime })R_{\mu }}{R^{\alpha }u_{\alpha }(t^{\prime })}%
=\right. \\
&&\left. =\frac{\frac{d}{ds}r_{\mu }(t)R_{k}-\frac{d}{ds}r_{k}(t)R_{\mu }}{%
R^{\alpha }u_{\alpha }(t)}.\right.  \notag
\end{eqnarray}%
Substituting this expression in Eq.(\ref{Eq.-----30}) there follows
\begin{eqnarray}
&&\overline{G}_{\mu }=2c\left( \frac{q}{c}\right) ^{2}\left[ \frac{u_{\mu
}(t^{\prime })}{R^{\alpha }u_{\alpha }(t^{\prime })R^{\beta }u_{\beta }(t)}%
\right. \\
&&\left. -\frac{R_{\mu }}{R^{\alpha }u_{\alpha }(t^{\prime })\left[ R^{\beta
}u_{\beta }(t)\right] ^{2}}u^{m}(t^{\prime })u_{m}(t)\right] _{t^{\prime
}=t-t_{ret}}.  \notag
\end{eqnarray}%
Invoking Lemma 3 this delivers Eq.(\ref{Eq.--------30bis}). Here we notice
that the proper-time derivatives $\frac{dr_{\mu }(t-t_{ret})}{ds}$ and $%
\frac{d^{2}r_{\mu }(t-t_{ret})}{ds^{2}}$ are evaluated invoking the chain
rule \ This is obtained by introducing the diffeomorphism $t\rightarrow
s(t)\equiv s$ [and similarly $t^{\prime }\rightarrow s^{\prime }(t^{\prime
})\equiv s^{\prime }$ ] with its inverse transformation $s\rightarrow t(s).$
It follows $t^{\prime }(s^{\prime })=t(s)-t_{ret}(t(s)),$ which proves\ that
Eq.(\ref{Eq.--------30bis}) delivers a parametric representation of $%
\overline{G}_{\mu }$ in terms of the local arc length $s$. \textbf{Q.E.D.}

Thus, remarkably, Eq.(\ref{Eq.--------30bis}) shows that, when parametrized
in terms of the local arc length $s,$ the 4-vector $\overline{G}_{\mu }$
depends - at most - on first-order derivatives (with respect to $s$) of the
4-position, i.e., is a function only of the 4-bi-vector $\widehat{R}^{\alpha
} $and of the derivatives $u_{k}(t)\equiv \frac{dr^{k}(t)}{ds}$ and $\frac{%
dr^{k}(t-t_{ret})}{ds}.$

\bigskip

\section{6 - Properties of $\overline{G}_{\protect\mu }$}

In this section we intend to investigate the main properties of the 4-vector
$\overline{G}_{\mu }$ (and hence of the RR equation given above). We intend
to show that they are fully consistent with the basic principles of
classical mechanics. In particular it is immediate to prove that $\overline{G%
}_{\mu }$ fulfills :

\begin{itemize}
\item Galilei's principle of inertia: in fact, in the case of inertial
motion it results identically $\overline{G}_{\mu }\equiv 0$;

\item the characteristic property of the Lorentz force, i.e., the Lorentzian
constraint%
\begin{equation}
\overline{G}_{\mu }u^{\mu }=0.
\end{equation}

\item Newton's principle of determinacy and Einstein's causality principle.
\end{itemize}

Finally, it can be shown that:

\begin{itemize}
\item $\overline{G}_{\mu }$ is defined also in the case of "sudden forces".
\end{itemize}

These results are summarized in the following theorem:\bigskip

\textbf{THM.3 - Properties of }$\overline{G}_{\mu }$

\emph{In validity of THM.1 and THM.2, the vector }$\overline{G}_{\mu }$\emph{%
\ fulfills the following properties}$:$

$\emph{T3}_{1}$\emph{) in case of inertial motion in a given proper-time
interval }$\left[ s_{1},s_{2}\right] ,$\emph{\ there results identically }$%
\overline{G}_{\mu }\equiv 0$\emph{;}

$\emph{T3}_{2}$\emph{) if F}$_{\mu \nu }^{(ext)}(r(s))\equiv 0$ $\forall s$
\emph{in a given proper-time interval }$\left[ s_{1},s_{2}\right] $\emph{\
and with respect to an inertial frame}$,$\emph{\ then there results
identically }$\overline{G}_{\mu }\equiv 0$ \emph{,}$\forall s\in $\emph{\ }$%
\left[ s_{1},s_{2}\right] $ \emph{(Galilei's inertia principle); }

$\emph{T3}_{3}$\emph{) }$\overline{G}_{\mu }$ \emph{satisfies the Lorentzian
constraint condition}%
\begin{equation}
\overline{G}_{\mu }u^{\mu }=0.  \label{Eq.-----31}
\end{equation}

\emph{Moreover, assuming that the RR equation (\ref{Eq.----29}), with (\ref%
{Eq.-----30}), admits smooth solutions in the proper-time interval }$\left[
s_{a},s_{b}\right] ,$ \emph{in such an interval:}

$\emph{T3}_{4}$\emph{) }$\overline{G}_{\mu }$ \emph{fulfills the Einstein's
causality principle, namely for any }$s\in \left[ s_{a},s_{b}\right] ,$ $%
r^{\mu }(s)$ \emph{depends only on the past history of} $r_{\mu }(s),$ \emph{%
i.e. }$\left\{ r^{\mu }(s^{\ast }),\forall s^{\ast }\leq s\right\} ;$

$\emph{T3}_{5}$\emph{) }$\overline{G}_{\mu }$ \emph{fulfills Newton's
determinacy principle, namely for any }$s_{o}\in \left[ s_{a},s_{b}\right] ,$
\emph{the knowledge of the particle initial state} $\left\{ r^{\mu
}(s_{o}),u_{\mu }(s_{o})\right\} $ \emph{determines uniquely the particle
state }$\left\{ r^{\mu }(s),u_{\mu }(s)\right\} $ \emph{at any }$s\geq s_{o}$%
\emph{\ which belongs to }$\left[ s_{a},s_{b}\right] ;$

$\emph{T3}_{6}$\emph{) }$\overline{G}_{\mu }$ \emph{is defined also in the
case of "sudden forces". For example, let us require that the external EM
field has the form}%
\begin{equation}
\overline{\emph{F}}_{\mu \nu }^{(ext)}(r(s))\equiv \left\{
\begin{array}{ccc}
0 &  & s\leq s_{0} \\
\emph{F}_{\mu \nu }^{(0)} &  & s>0%
\end{array}%
\right.
\end{equation}%
\emph{with }$F_{\mu \nu }^{(0)}$\emph{\ a constant 4-tensor and }$s_{0}$%
\emph{\ }$\in \left[ s_{a},s_{b}\right] $\emph{. In such a case one can
prove that the solution of the RR equation exists and is unique. }

\emph{Proof} - To prove propositions $T3_{1}$ and $T3_{2}$ let us assume
that in the interval $\left[ s_{1},s_{2}\right] $ the motion is inertial,
namely that $\frac{d}{ds}u_{\mu }\equiv 0$\emph{,}$\forall s$ $\in \left[
s_{1},s_{2}\right] .$ This implies, that in $\left[ s_{1},s_{2}\right] ,$ $%
u_{\mu }\equiv u_{0\mu },$ with $u_{0\mu }$ denoting a constant 4-vector
velocity. It follows $\forall s,s^{\prime }\in $ $\left[ s_{1},s_{2}\right]
, $ $r_{\mu }(s)=r_{\mu }(s^{\prime })+u_{0\mu }(s^{\prime })(s-s^{\prime })$
and $R_{\mu }=u_{0\mu }(s)(s-s^{\prime }).$ Hence, there results identically
\begin{eqnarray}
&&\left. \frac{\frac{dr_{\mu }(t-t_{ret})}{ds}R_{k}-\frac{dr_{k}(t-t_{ret})}{%
ds}R_{\mu }}{R^{\alpha }u_{\alpha }(s)}=\right. \\
&&\left. =\frac{u_{0\mu }u_{0k}(s)(s-s^{\prime })-u_{0\mu
}u_{0k}(s)(s-s^{\prime })}{s-s^{\prime }}\equiv 0.\right.  \notag
\end{eqnarray}%
Propositions $T3_{3},\emph{T3}_{4}$ and $\emph{T3}_{5}$ follow, similarly,
by direct inspection of Eqs.(\ref{Eq.----29}) and (\ref{Eq.-----30}), or
similarly Eq.(\ref{Eq.--------30bis}). In particular, $\emph{T3}_{5}$ is an
immediate consequence of THM.1 and the fact that the RR equation defines a
well-posed initial-value problem. Finally, the proof of proposition $\emph{T3%
}_{6}$ can be obtained by explicit construction of the solution of the RR
equation (see analogous treatment given in Ref.\cite{Dorigo2008a} for the
weakly-relativistic LAD equation). \ \textbf{Q.E.D.}

\section{7 - Non-existence of the point-charge limit}

An important aspect of the present formulation concerns the validity of the
RR equation obtained letting
\begin{equation}
\sigma \rightarrow 0^{+}  \label{point-particle}
\end{equation}%
(\emph{point-charge limit}) in the definition of $\overline{G}_{\mu }$ [see
Eq.(\ref{Eq.-----30}) or (\ref{Eq.--------30bis})].\ Here we intend to prove
that:

\begin{itemize}
\item \emph{the exact RR equation is not defined in the limit} (\ref%
{point-particle}) [see following THM.4]. In other words, \emph{the
point-charge limit [for }$\overline{G}_{\mu }$\emph{]} \emph{is not defined}.
\end{itemize}

To establish the result let us introduce yet another representation of the
4-vector $\overline{G}_{\mu }$ which makes explicit its dependence in terms
of the parameter $\sigma ,$ the radius of the spherical charge distribution.
For definiteness let us introduce the position%
\begin{equation}
\mathbf{w\equiv v}(t^{\prime })+\frac{1}{(t-t^{\prime })}\int\limits_{t^{%
\prime }}^{t}dt_{2}\mathbf{a}(t_{2})(t-t_{2}).  \label{Eq.-alfa-1}
\end{equation}%
Eq.(\ref{Eq.-----24}) can also be written as
\begin{equation}
R^{\alpha }R_{\alpha }=c^{2}(t-t^{\prime })^{2}\left\{ 1-\frac{w^{2}}{c^{2}}%
\right\} =\sigma ^{2},  \label{Eq.-alfa-2}
\end{equation}%
so that the delay-time $t^{\prime }-t=t_{ret},$ with $t_{ret}>0,$ \ reads%
\begin{equation}
t_{ret}=\frac{\sigma }{c\sqrt{1-\frac{w^{2}}{c^{2}}}}.  \label{Eq.-alfa-3}
\end{equation}%
Here it is obvious that for all $\sigma >0$ the following inequalities must
hold
\begin{equation}
\left\{
\begin{array}{c}
1-w^{2}/c^{2}>0, \\
t_{ret}=\frac{\sigma }{c\sqrt{1-\frac{w^{2}}{c^{2}}}}>\frac{\sigma }{c}, \\
R^{\alpha }v_{\alpha }(s)=\frac{c\sigma }{\sqrt{1-\frac{w^{2}}{c^{2}}}}\left[
1-\frac{1}{c^{2}}\frac{d\mathbf{r}(t)}{dt}\cdot \mathbf{w}\right] >0, \\
1-\frac{1}{c^{2}}\frac{d\mathbf{r}(t)}{dt}\cdot \mathbf{w}>0\mathbf{.}%
\end{array}%
\right.  \label{Eq.-alfa-4}
\end{equation}%
Thus, introducing \ the 4-vector $X_{k}\equiv \frac{1}{c\sigma }\sqrt{1-%
\frac{w^{2}}{c^{2}}}R_{k}=\left\{ c,\mathbf{w}\right\} $ one obtains for $%
\overline{G}_{\mu }$ the representation%
\begin{eqnarray}
&&\left. \overline{G}_{\mu }=-2c\left( \frac{q}{c}\right) ^{2}\frac{\sqrt{1-%
\frac{w^{2}}{c^{2}}}u^{k}(s)}{\sigma c\left[ 1-\frac{1}{c^{2}}\frac{d\mathbf{%
r}(t)}{dt}\cdot \mathbf{w}\right] }\right.  \label{Eq.-alfa-5} \\
&&\frac{d}{ds}\left\{ \frac{\frac{dr_{\mu }(t-t_{ret})}{ds}X_{k}-\frac{%
dr_{k}(t-t_{ret})}{ds}X_{\mu }}{\left( 1-\frac{1}{c^{2}}\frac{d\mathbf{r}(t)%
}{dt}\cdot \mathbf{w}\right) }\sqrt{1-\frac{\mathbf{v}^{2}(t)}{c^{2}}}%
\right\} ,  \notag
\end{eqnarray}%
which displays explicitly its dependence in terms of $\sigma .$

\subsection{The singular limit $\protect\sigma \rightarrow 0^{+}$}

Let us now investigate the limit $\sigma \rightarrow 0^{+}$ for $\overline{G}%
_{\mu }.$ The following (non-existence) theorem holds:\bigskip

\textbf{THM.4 - Non-existence of the point-charge limit for }$\overline{G}%
_{\mu }$

\emph{The limit }$\lim_{\sigma \rightarrow 0^{+}}\overline{G}_{\mu }$\emph{\
is not defined. In other words: for spherically symmetric charges the RR
4-vector is not defined in the limit (\ref{point-particle}).}

\emph{Proof} - Let us introduce the absurd hypothesis that the following
limits exist:%
\begin{equation}
\lim_{\sigma \rightarrow 0^{+}}\left( 1-\frac{w^{2}}{c^{2}}\right) >0,
\label{hyp-a}
\end{equation}%
\begin{equation}
\lim_{\sigma \rightarrow 0^{+}}\left( 1-\frac{1}{c^{2}}\frac{d\mathbf{r}(t)}{%
dt}\cdot \mathbf{w}\right) >0,  \label{hyp-b}
\end{equation}%
and moreover that for non-inertial motion there results%
\begin{equation}
0<\lim_{\sigma \rightarrow 0^{+}}\left\vert \frac{d}{ds}\widehat{H}_{\mu
k}\right\vert <\infty ,  \label{hyp-c}
\end{equation}%
where
\begin{eqnarray}
\widehat{H}_{\mu k} &\equiv &\frac{\frac{dr_{\mu }(t-t_{ret})}{ds}X_{k}-%
\frac{dr_{k}(t-t_{ret})}{ds}X_{\mu }}{\left( 1-\frac{1}{c^{2}}\frac{d\mathbf{%
r}(t)}{dt}\cdot \mathbf{w}\right) } \\
&&\sqrt{1-\frac{\mathbf{v}^{2}(t)}{c^{2}}.}  \notag
\end{eqnarray}%
In such a case, invoking Eq.(\ref{Eq.-alfa-5}) for $\overline{G}_{\mu }$, it
follows necessarily
\begin{equation}
\lim_{\sigma \rightarrow 0^{+}}\overline{G}_{\mu }\varpropto \lim_{\sigma
\rightarrow 0^{+}}\frac{2}{c^{2}\sigma ^{2}}=\infty .
\end{equation}%
Hence, in validity of (\ref{hyp-a})-(\ref{hyp-c}) the limit $\lim_{\sigma
\rightarrow 0^{+}}\overline{G}_{\mu }$ does not exist. On the other hand if
one of the inequalities (\ref{hyp-a})-(\ref{hyp-c}) is violated, the motion
defined by the RR equation [Eq.(\ref{Eq.----29})] is non-physical, which
brings again the same conclusion. \textbf{Q.E.D.}

\section{8 - Short-time approximation and the LAD equation}

A crucial point in the Dirac evaluation of the LAD equation \cite{Dirac1938}
was the power-series expansion of the retarded potential in terms of a
suitably defined small dimensionless parameter $\xi ,$ related to the
proper-time difference between emission ($t^{\prime }$) and observation ($t$%
) times,
\begin{equation}
0<\xi \equiv \frac{(t-t^{\prime })}{t},  \label{ordering -1}
\end{equation}%
to be assumed as infinitesimal (\emph{short-time ordering}). The same
approach was also adopted by DeWitt and Brehme \cite{DeWitt1960} in their
covariant generalization of the LAD equation valid in curved space-time.

In analogy, here we introduce a power-series expansion with respect to the
dimensionless parameter $\xi $ of the form%
\begin{equation}
\overline{G}_{\mu }=\sum\limits_{k=0}^{\infty }\xi ^{k}\overline{G}_{\mu
}^{(k)},  \label{xsi-series}
\end{equation}%
which is assumed to converge for
\begin{equation}
\xi \ll 1  \label{asymptotic ordering}
\end{equation}%
(\emph{short-time asymptotic ordering}). The power series expansion is
actually obtained by introducing a Taylor expansion for the 4-position
vector $r^{\mu }(t-t_{ret})$ in terms of the retarded time $t^{\prime },$
namely letting
\begin{equation}
r^{\mu }(t-t_{ret})=\sum\limits_{k=0}^{\infty }\frac{(t^{\prime }-t)^{k}}{k!}%
\frac{d^{k}r^{\mu }(t)}{dt^{k}}.  \label{Taylor-series}
\end{equation}%
Manifestly, for the validity (i.e., the convergence) of the series, a
prerequisite is that $r^{\mu }(s)\equiv r^{\mu }(t(s))$ is a $C^{(\infty )}$
function. In turn, this requires that also the Faraday tensor generated by
the external EM field, $F_{\mu }^{(ext)\nu },$ must be $C^{(\infty )}.$ The
use of the expansion (\ref{ordering -1}) to represent the 4-vector $%
\overline{G}_{\mu }$ reduces, formally, the RR equation to a local and
infinite-order ordinary differential equation. In view of THM.1 and the
assumed convergence of the series its (infinitely) smooth solution must
still exist and be uniquely defined. As a side consequence, this means that
the initial conditions for such an equation must necessarily be considered
as uniquely prescribed in terms of the initial conditions defined above [see
Eqs.(\ref{initial conditions-1}) and (\ref{initial conditions-2})] and the
same RR-equation. Nevertheless, despite these features, the full
series-representation of the RR equation obtained in this way appears
practically useless for actual applications.

As an alternative, however, \ assuming $\xi $ as infinitesimal an asymptotic
approximation for $\overline{G}_{\mu }$ [and the exact RR equation\ Eq.(\ref%
{Eq.---------------20B}) or equivalent Eq.(\ref{Eq.----29})] can in
principle be achieved, subject again to suitable smoothness assumptions to
be imposed on the external field. Here we intend to prove, in particular,
that in this way:

\begin{itemize}
\item \emph{the relativistic LAD equation is recovered as a leading-order
asymptotic approximation to the exact RR equation}. In fact, provided
suitable smoothness conditions are met by the external field, the 4-vector $%
\overline{G}_{\mu }$ recovers\ \emph{asymptotically }- in a suitable
approximation - the usual form of RR equation provided by the LAD equation.
This conclusion is achieved by introducing for the 4-vector $\overline{G}%
_{\mu }$ an asymptotic expansion with respect to the dimensionless parameter
$\xi \ll 1,$ obtained by means of a truncated Taylor expansion in terms of
the retarded time $t^{\prime },$ i.e., of the form%
\begin{equation}
r^{\mu }(t-t_{ret})=\sum\limits_{k=0}^{N}\frac{(t^{\prime }-t)^{k}}{k!}\frac{%
d^{k}r^{\mu }(t)}{dt^{k}},
\end{equation}%
with $N>0$ to be suitably prescribed. \
\end{itemize}

In such a case the following result holds:\bigskip

\textbf{THM.5 - First-order, short-time asymptotic approximation for }$%
\overline{G}_{\mu }$

\emph{Let us now assume that the EM-4-potential of the external field }$%
A_{\mu }^{(ext)}(r)$ \emph{is a smooth function of} $r.$ \emph{In such a
case, in validity of the asymptotic ordering (\ref{asymptotic ordering}) and
neglecting corrections of order }$\xi ^{N},$ \emph{with} $N\geq 1$ \emph{%
(first-order approximation})$,$\emph{\ the following asymptotic
approximation holds for }$\overline{G}_{\mu }$%
\begin{equation}
\left. \overline{G}_{\mu }\cong \left\{ m_{oEM}c\frac{d}{ds}u_{\mu }+g_{\mu
}\right\} \left[ 1+O(\xi )\right] \right. ,  \label{Eq.-beta}
\end{equation}%
\emph{with }$g_{\mu }$\emph{\ denoting the 4-vector}%
\begin{equation}
g_{\mu }=\frac{2}{3}\frac{q^{2}}{c}\left[ \frac{d^{2}}{ds^{2}}u_{\mu
}-u_{\mu }(s)u^{k}(s)\frac{d^{2}}{ds^{2}}u_{k}\right] ,  \label{Eq.-beta-0}
\end{equation}%
\emph{and }%
\begin{equation}
m_{oEM}\equiv \frac{q^{2}}{c^{2}\sigma }\frac{1}{\left[ 1+\frac{(t-t^{\prime
})}{2}\frac{d}{ds}\frac{1}{\gamma }\right] ^{2}}  \label{Eq.-beta-00}
\end{equation}%
\emph{\ the EM mass.}

\emph{Proof} - To reach the proof let us first evaluate asymptotic
expansions for the 4-vectors $R^{k},$ $\frac{dr_{\mu }(t-t_{ret})}{ds},$ the
4-scalar $R^{\alpha }u_{\alpha }(s)$ and the time delay $t-t^{\prime }\equiv
t_{ret}.$ Neglecting corrections of order $\xi ^{N}$ with $N>3,$ and
denoting $\gamma \equiv \gamma (t(s))\equiv 1/\sqrt{\left(
1-v^{2}(t(s)\right) /c^{2}}$ and $u^{k}\equiv u^{k}(t(s)),$ one obtains by
Taylor expansion

\begin{eqnarray}
&&R^{k}\cong \frac{c(t-t^{\prime })}{\gamma }u^{k}-\frac{c^{2}(t-t^{\prime
})^{2}}{2\gamma }\frac{d}{ds}\left( \frac{u^{k}}{\gamma }\right) +
\label{Eq.-beta-1} \\
&&+\frac{c^{3}(t-t^{\prime })^{3}}{6\gamma }\frac{d}{ds}\left( \frac{1}{%
\gamma }\frac{d}{ds}\frac{u^{k}}{\gamma }\right)  \notag
\end{eqnarray}%
and similarly denoting $r_{\mu }\equiv r_{\mu }(t(s)),$
\begin{eqnarray}
&&\left. \frac{dr_{\mu }(t-t_{ret})}{ds}\cong \right.  \label{Eq.-beta-2} \\
&\cong &\frac{dr_{\mu }}{ds}-\frac{c(t-t^{\prime })}{\gamma }\frac{%
d^{2}r_{\mu }}{ds^{2}}+\frac{c^{2}(t-t^{\prime })^{2}}{2\gamma ^{2}}\frac{%
d^{3}r_{\mu }}{ds^{3}}.  \notag
\end{eqnarray}%
Thus, Eqs.(\ref{Eq.-beta-1}) and (\ref{Eq.-beta-2}) imply%
\begin{eqnarray}
&&R^{\alpha }u_{\alpha }(s)\cong \frac{c(t-t^{\prime })}{\gamma }-\frac{%
c^{2}(t-t^{\prime })^{2}}{2\gamma }\frac{d}{ds}\frac{1}{\gamma }+
\label{Eq.-beta-3} \\
&&\left. +\frac{c^{3}(t-t^{\prime })^{3}}{6\gamma }\frac{d}{ds}\left( \frac{1%
}{\gamma }\frac{d}{ds}\frac{1}{\gamma }u^{\alpha }\right) u_{\alpha }\right.
,  \notag
\end{eqnarray}%
where $\frac{d}{ds}\frac{1}{\gamma }=\frac{d}{ds}\sqrt{1-\frac{v^{2}}{c^{2}}}%
=-\gamma \frac{\mathbf{v}(t)\cdot \mathbf{a}(t)}{c^{2}}.$ Finally, we notice
that there results
\begin{eqnarray}
&&\left. t-t^{\prime }=\frac{\sigma }{c\sqrt{1-\frac{w^{2}}{c^{2}}}}\cong
\right.  \label{Eq.-beta-4} \\
&\cong &\frac{\sigma \gamma }{c}\left[ 1+\frac{3}{2}\frac{\sigma \gamma }{c}%
\frac{\mathbf{v}(t)\cdot \mathbf{a}(t)}{v(t)^{2}}\right]  \notag
\end{eqnarray}%
\bigskip \bigskip

By substituting Eqs.(\ref{Eq.-beta-1})-(\ref{Eq.-beta-4}) in Eq.(\ref%
{Eq.-----30}) [or equivalent in Eq.(\ref{Eq.--------30bis})] it is immediate
to recover after straightforward calculations Eq.(\ref{Eq.-beta}). \textbf{%
Q.E.D.}

We remark that Eq.(\ref{Eq.-beta-0}) for $g_{\mu }$ coincides formally with
the usual expression of the EM self-force adopted in the LAD equation (see
related discussion in Ref.\cite{Dorigo2008a}). $\ $However, to recover the
customary expression of the EM mass usually given [for the LAD equation]
(see for example Ref.\cite{Rohrlich2001}), requires retaining only the
leading-order approximation
\begin{equation}
m_{oEM}\cong \frac{q^{2}}{c^{2}\sigma }\left[ 1+O(\xi )\right] ,
\label{Eq.RR-02}
\end{equation}%
This amounts to ignore the correction factor%
\begin{equation}
\frac{1}{\left[ 1+\frac{(t-t^{\prime })}{2}\frac{d}{ds}\frac{1}{\gamma }%
\right] ^{2}}\cong 1+\frac{\sigma }{c}\frac{\mathbf{v}(t)\cdot \mathbf{a}(t)%
}{c^{2}},
\end{equation}%
i.e., a term of order $\xi ^{0}$ in Eq.(\ref{Eq.-beta-00}).\ Hence this
approximation is not sufficient, since the term $g_{\mu }$ in Eq.(\ref%
{Eq.-beta-0}) is of order $\xi ^{0}$ too. We conclude that, for consistency,
in place of (\ref{Eq.RR-02}), the more accurate approximation (\ref%
{Eq.-beta-00}) should be used for the EM mass $m_{oEM}$.

An important issue is related to the conditions of smoothness - required by
THM.5 for the validity of Eqs.(\ref{Eq.-beta})-(\ref{Eq.-beta-00}) - which
must be imposed on the external EM field, i.e., on $A_{\mu }^{(ext)}(r(s)).$
It is obvious, in particular, that locally discontinuous (in $s$) external
fields \emph{must generally be excluded}, since the previous expansions [see
Eqs.(\ref{Eq.-beta-1})-(\ref{Eq.-beta-4})] manifestly do not hold near the
discontinuities. An example is provided by so-called "sudden forces". These
occur when the corresponding Faraday tensor $F_{\mu \nu }^{(ext)}(r(s))$ is
permitted to be locally discontinuous with respect to $s$ (which may be
achieved by turning on and off repeatedly the external EM field). For the
validity of THM.5 this case must generally be excluded. In fact, it is
obvious that the\ Taylor expansions (\ref{Eq.-beta-1})-(\ref{Eq.-beta-4})
generally do not hold in the neighborhood of the discontinuities. This
clearly prevents also the validity of the LL equation as well of analogous
asymptotic approximation of Eq.(\ref{Eq.-beta}) (see also related discussion
in Ref.\cite{Dorigo2008a}).

\section{9 - Weakly-relativistic approximation}

Although the covariant representation given by Eq.(\ref{Eq.--------30bis})
is of general validity, it is worth discussing here also its
weakly-relativistic approximation. This enables a direct comparison with the
original Lorentz approach \cite{Lorentz} and the known result obtained by
Sommerfeld, Page, Caldirola and Yaghjian \cite%
{Sommerfeld,Page,Caldirola,Yaghjian} in the case of a finite-size
spherical-shell charge, a fact which is relevant not merely for historical
reasons.\ Indeed, as previously pointed out \cite{Dorigo2008a}, also the
weakly-relativistic LAD equation exhibits the same difficulties
characteristic of the relativistic LAD equation. In particular, it yields a
third-order ordinary differential equation which exhibits the known physical
inconsistencies (violation of NPD and GPI, existence of runaway solutions
which blow up in time, etc.). Here we intend to show how, even in the
weakly-relativistic approximation, the present theory is able to overcome
such difficulties. For the sake of definiteness, let us determine the
asymptotic approximation for $\overline{G}_{\mu },$ obtained by assuming
\begin{equation}
\beta \equiv v(t)/c\ll 1  \label{beta}
\end{equation}%
(\emph{weakly-relativistic approximation}). For this purpose let us
introduce a Taylor expansion with respect to $\beta ,$ while leaving
unchanged the dependence in terms of the retarded time $t^{\prime }$. As
shown in Appendix E, in such a case the following result holds:\bigskip

\textbf{THM.6 - Weakly-relativistic asymptotic approximation for }$\overline{%
G}_{\mu }$

\emph{In validity of the asymptotic ordering (\ref{beta}) and neglecting
corrections of order }$\beta ^{n},$ \emph{with} $n\geq 3,$\emph{\ the
following asymptotic approximation holds for }$\overline{G}_{\mu }:$%
\begin{equation}
\left. \overline{G}_{\mu }\cong \left( G_{0}=0,\mathbf{G}\right) \right. ,
\end{equation}%
\emph{where:}

$T6_{1})${\footnotesize \ }\emph{first asymptotic approximation: in the case
of the representation (\ref{Eq.-----30}) the 3-vector} $\mathbf{G}$ \emph{%
reads:} \emph{\ }%
\begin{eqnarray}
&&\mathbf{G}\cong -\frac{2}{\sigma }\left( \frac{q}{c}\right) ^{2}\left[
\frac{d}{dt}\mathbf{v}(t-\frac{\sigma }{c})+\right.  \label{Eq.50} \\
&&\left. \left. +\frac{c}{\sigma }\mathbf{v(}t-\frac{\sigma }{c})-\frac{c^{2}%
}{\sigma ^{2}}\left\{ \mathbf{r}(t)\mathbf{-r}(t-\frac{\sigma }{c})\right\} %
\right] \right. ;  \notag
\end{eqnarray}%
$T6_{2})$\emph{\ second asymptotic approximation: in the case of the
representation (\ref{Eq.--------30bis}), instead, the 3-vector} $\mathbf{G}$
\emph{becomes: }%
\begin{eqnarray}
&&\mathbf{G}\cong 2c\left( \frac{q}{c}\right) ^{2}\frac{1}{\sigma ^{2}}\left[
\frac{d\mathbf{r}(t-t_{ret})}{dt}-\right.  \label{Eq.51} \\
&&\left. -\frac{\mathbf{r}(t)\mathbf{-r}(t-t_{ret})}{\left\vert (t-t^{\prime
})\right\vert }\right] ;  \notag
\end{eqnarray}

$T6_{3})$ \emph{finally, upon invoking also the short-time ordering (\ref%
{ordering -1}) and a suitable condition of smoothness for the external EM
field, one recovers in both cases [Eqs.(\ref{Eq.50}) or (\ref{Eq.51})] the
usual weakly relativistic approximation:}%
\begin{equation}
\mathbf{G\cong g}+m_{EM}\overset{\cdot \cdot }{\mathbf{r}}(t)\mathbf{,}
\label{weakly-relativistic RR force}
\end{equation}%
\emph{where }%
\begin{eqnarray}
&&\left. \mathbf{g}\equiv -\frac{\mathbf{2}q^{2}}{3c^{3}}\overset{\cdot
\cdot \cdot }{\mathbf{r}},\right.  \label{NR self-force} \\
&&\left. m_{EM}\equiv \frac{q^{2}}{c^{2}\sigma },\right.  \label{EM mass}
\end{eqnarray}%
\emph{are respectively the well-known weakly-relativistic\ EM self-force
3-vector and the EM mass.}

\emph{Proof} \ (see Appendix E). We notice that the apparent non-uniqueness
of the two representations given above [Eqs.(\ref{Eq.50}) and (\ref{Eq.51})]
can be resolved by noting that the $\beta -$expansion should be actually
carried out also in terms of the delay-time $t_{ret}$ (which should be
considered itself of order $\beta ^{\alpha },$ with $\alpha >0$ to be
suitably defined)$.$ Indeed, if the short-time expansion is introduced, as
found in Appendix E, Eqs.(\ref{weakly-relativistic RR force}) both imply
Eqs.(\ref{NR self-force}) and (\ref{EM mass}). \ In the same sense, Eqs.(\ref%
{Eq.50}) and (\ref{Eq.51}) can also be proven to be in agreement with the
well-known Sommerfeld-Page-Caldirola-Yaghjian result \cite%
{Sommerfeld,Page,Caldirola,Yaghjian} for weakly-relativistic spherical-shell
charges. The resulting equations, (\ref{NR self-force}) and (\ref{EM mass}),
are manifestly consistent with the customary weakly-relativistic
approximation for the LAD equation (see, for example, also related
discussion in Ref.\cite{Dorigo2008a}). \

\section{10 - Concluding remarks}

In this paper an exact solution has been obtained for the RR problem. The
result has been achieved in the case of a spherical-shell finite-size
charge.\ As a main consequence, the exact RR equation, describing the
relativistic dynamics of such a particle in the presence of its EM
self-field has been achieved (see THM.1 and THM.2). Although its charge has
been assumed as spatially distributed, we have shown that, by assuming the
mass as point-wise localized, the dynamics is reduced to that of a point
particle. The resulting RR equation appears free from all the difficulties
met by previously classical RR equations (THM.1-THM.3). \ In particular,
besides being fully relativistic, the new equation:

1) \textit{has been achieved via a variational formulation based on the
adoption of the Hamilton variational principle}. The treatment has been made
transparent by adopting a synchronous form of the variational principle;

2) \textit{unlike the LAD equation}: results consistent with the\ Newton's
principle of determinacy, Einstein principle of causality, Galilei law of
inertia and does not exhibit so-called runaway solutions;

3) \textit{unlike} \textit{the LL equation}: \ does not involve the adoption
of iterative approaches for its derivation;

4) \textit{unlike the LAD and LL equations}: is valid also in the case of
sudden forces and does not exhibit any singular behavior (i.e., provided the
radius of the charge $\sigma $ remains strictly positive);

5) \textit{unlike} \textit{all previous equations} (\textit{LAD and LL \ and
the Medina equations):} it is not asymptotic.

6) \textit{unlike} \textit{in the Medina approach}: the variational approach
is based on the Hamilton variational principle in the ordinary phase-space,
which allows us to retain the customary formulation of classical mechanics
and classical electrodynamics.

In addition, as a side result, we have pointed out a correction to the LAD
equation, appearing in the EM mass, which is demanded by the perturbative
expansion [see Eq.(\ref{Eq.-beta-00})].

The theory developed in this paper has, potentially, deep and wide-ranging
implications. These are related, in particular, to the description of
relativistic dynamics of systems of classical finite-size charged particles.
The conceptual simplicity of the present approach and its general
applicability to arbitrary systems of charges of this type make the present
results of extraordinary relevance for relativistic theories (such as
kinetic theory of charged particles and gyrokinetic theory for
magnetoplasmas) and related applications in astro- and plasma physics.

\section*{Acknowledgments}

Useful comments by A. Beklemishev (Budker Institute of Nuclear Physics,
Novosibirsk, Russia Federation), J. Miller (International School for
Advanced Studies, Miramare, Trieste, Italy and Oxford University, Oxford,
UK) and P. Nicolini (Department of Mathematics and Informatics, Trieste
University, Italy) are acknowledged. This work has been developed in
cooperation with the CMFD Team, Consortium for Magnetofluid Dynamics
(Trieste University, Trieste, Italy), within the framework of the MIUR
(Italian Ministry of University and Research) PRIN Programme: \textit{%
Modelli della teoria cinetica matematica nello studio dei sistemi complessi
nelle scienze applicate}. Support is acknowledged from GNFM (National Group
of Mathematical Physics) of INDAM (Italian National Institute for Advanced
Mathematics).

\section{APPENDIX\ A: surface-average operator}

Following the notations introduced in Sec.4b and in case of flat space-time,
if $A(r+\sigma n)$\ is a smooth (tensor) function of the 4-position vector $%
r+\sigma n,$ we define its surface-average as%
\begin{equation}
\overline{A}(r)=\frac{1}{4\pi }\int d\Sigma (n)A(r+\sigma n).
\label{surface-average}
\end{equation}
In particular, identifying $A$ with the Faraday tensor $F_{\mu }^{\nu
}(r+\sigma n),$ its surface-average is\emph{\ }%
\begin{equation}
\overline{F}_{\mu }^{\nu }(r)=\frac{1}{4\pi }\int d\Sigma (n)F_{\mu }^{\nu
}(r+\sigma n).  \label{surface average -2}
\end{equation}%
\emph{\ }

\section{APPENDIX\ B: integral representation for $A_{\protect\mu }^{(self)}$
(case of a point charge)}

The integral representation (\ref{integral representation}) for $A_{\mu
}^{self}$ can also be obtained directly from Maxwell's equations$.$ Let us
consider first the case of a point-charge. By assumption $A_{\mu }^{self}$
satisfies Maxwell's equations (in flat space-time)
\begin{equation}
\partial _{\mu }F^{\mu \nu (self)}=\frac{4\pi }{c}j^{\nu },  \label{6}
\end{equation}%
where for a point particle:%
\begin{eqnarray}
j^{\mu }(r^{\nu }) &=&q\int ds^{\prime }u^{\mu }(s^{\prime })  \label{4} \\
&&\delta ^{(4)}\left( r-r(s^{\prime })\right)
\end{eqnarray}%
(with $\delta ^{(4)}\left( r-r(\tau )\right) $ denoting the 4-dimensional
Dirac delta). There results therefore
\begin{equation}
A^{\mu (self)}=\frac{4\pi }{c}\int d^{4}r^{\prime }G(r-r^{\prime })j^{\mu
}(r^{\prime }),  \label{A3}
\end{equation}%
where $G(r-r)$ is the retarded Green function which satisfies the equation%
\begin{equation}
\boxdot G(r-r^{\prime })=\delta ^{(4)}\left( r-r^{\prime }\right)
\end{equation}%
and is such that
\begin{equation}
G(r-r^{\prime })=0
\end{equation}%
for $r^{0}<r^{\prime 0}.$ It follows%
\begin{equation}
G(r-r^{\prime })=\frac{1}{2\pi }\delta (R^{\mu }R_{\mu })\Theta
(r^{0}-r^{\prime 0}),
\end{equation}%
and hence%
\begin{eqnarray}
&&\left. A^{\mu (self)}(r)=\frac{4\pi }{c}\int d^{4}r^{\prime }\frac{1}{2\pi
}\delta (R^{\mu }R_{\mu })\right.  \notag \\
&&\Theta (r^{0}-r^{\prime 0}) \\
&&\left. q\int ds^{\prime }u^{\mu }(s^{\prime })\delta ^{(4)}\left(
r^{\prime }-r(s^{\prime })\right) ,\right.  \notag
\end{eqnarray}%
namely%
\begin{eqnarray}
&&\left. A_{\mu }^{(self)}(r)=\frac{2}{c}\int d^{4}r^{\prime }\delta (R^{\mu
}R_{\mu })\Theta (r^{0}-r^{\prime 0})q\right. \\
&&\left. \int ds^{\prime }u_{\mu }(s^{\prime })\delta ^{(4)}\left( r^{\prime
}-r(s^{\prime })\right) .\right.  \notag
\end{eqnarray}%
This implies also%
\begin{equation}
A_{\mu }^{(self)}(r)=\frac{2q}{c}\int ds^{\prime }u_{\mu }(s^{\prime
})\delta (R^{\mu }(s^{\prime })R_{\mu }(s^{\prime })),
\end{equation}%
where $R^{\mu }(s^{\prime })=r^{\mu }-r^{\mu }(s^{\prime }).$ The last
integral can also be written as%
\begin{equation}
A_{\mu }^{(self)}(r)=\frac{2q}{c}\int dr_{\mu }^{\prime }\delta (R^{\mu
}R_{\mu }).  \label{10}
\end{equation}%
This is an integral representation for $A_{\mu }^{(self)}(r),$ by
construction equivalent to Eq.(\ref{A3}).

\section{APPENDIX\ C: integral representation for $A_{\protect\mu }^{(self)}$
(case of a spherical-shell charge)}

To prove that the differential and integral representations for $\overline{A}%
_{\mu }^{self}$ (\ref{differential representation}) and (\ref{integral
representation}) are equivalent it is sufficient to notice that the
following identity holds:%
\begin{eqnarray}
&&\left. \delta (R^{\alpha }R_{\alpha }-\sigma ^{2})=\right.  \label{C-1} \\
&&\left. =\delta (t-t^{\prime }-t_{ret})\right.  \notag \\
&&\frac{1}{2c^{2}\left\vert (t-t^{\prime })-\frac{1}{c^{2}}\frac{d\mathbf{r}%
(t^{\prime })}{dt^{\prime }}\cdot (\mathbf{r-r}^{\prime })\right\vert }.
\notag
\end{eqnarray}%
In fact there follows
\begin{eqnarray}
&&\left. \overline{A}_{\mu }^{(self)}(r)=\frac{2q}{c}\int_{_{1}}^{2}dr_{\mu
}^{\prime }\delta (R^{\alpha }R_{\alpha }-\sigma ^{2})=\right. \\
&=&\frac{2q}{c}\left[ \frac{1}{2c^{2}\left\vert (t-t^{\prime })-\frac{1}{%
c^{2}}\frac{d\mathbf{r}(t^{\prime })}{dt^{\prime }}\cdot (\mathbf{r-r}%
^{\prime })\right\vert }\frac{dr_{\mu }^{\prime }}{dt^{\prime }}\right]
_{t^{\prime }=t-t_{ret}}=  \notag \\
&=&\frac{q}{c}\left[ \frac{u_{\mu }(t^{\prime })}{R^{\alpha }u_{\alpha
}(t^{\prime })}\right] _{t^{\prime }=t-t_{ret}},  \notag
\end{eqnarray}%
which recovers immediately Eq.(\ref{differential representation}).

\section{Appendix D - Other lemmas}

\textbf{LEMMA 2 - Synchronous variation of }$\Delta S_{1}$

\emph{The synchronous variation of}\textbf{\ }$\Delta S_{1}(r^{\mu })$ \emph{%
reads }%
\begin{equation}
\delta \Delta S_{1}=\delta A+\delta B,  \label{Eq.---A.3}
\end{equation}%
\emph{where}%
\begin{equation}
\begin{array}{c}
\delta A\equiv -4\left( \frac{q}{c}\right) ^{2}g_{\mu \nu }\frac{1}{4\pi }
\\
\int d\Sigma (\mathbf{n})\int_{1}^{2}\delta r^{\mu }d\left[
\int_{1}^{2}dr^{\prime \nu }\delta (R^{\alpha }R_{\alpha }-\sigma ^{2})%
\right] , \\
\delta B\equiv 4\left( \frac{q}{c}\right) ^{2}g_{\alpha \beta }\frac{1}{4\pi
} \\
\int d\Sigma (\mathbf{n})\int_{1}^{2}dr^{\prime \beta
}\int_{1}^{2}dr^{\alpha }\delta r^{\mu }\frac{\partial }{\partial r^{\mu }}%
\delta (R^{k}R_{k}-\sigma ^{2}).%
\end{array}
\label{Eq.---A.4}
\end{equation}%
\emph{There results respectively:}%
\begin{eqnarray}
&&\left. \delta A\equiv 4\left( \frac{q}{c}\right) ^{2}g_{\mu \nu }\frac{c}{%
4\pi }\int d\Sigma (\mathbf{n})\right.  \label{Eq.---A.5-} \\
&&\int_{1}^{2}\delta r^{\mu }dr^{k}\left[ A_{k}^{\nu }\right] _{t^{\prime
}=t-t_{ret}},  \notag
\end{eqnarray}

\begin{eqnarray}
&&\left. \delta B\equiv -4\left( \frac{q}{c}\right) ^{2}g_{\alpha \beta }%
\frac{c}{4\pi }\int d\Sigma (\mathbf{n})\right.  \label{Eq.---A.6-} \\
&&\int_{1}^{2}dr^{\alpha }\delta r^{\mu }\left[ A_{\mu }^{\beta }\right]
_{t^{\prime }=t-t_{ret}},  \notag
\end{eqnarray}%
where
\begin{eqnarray}
&&\left. A_{k}^{\nu }\equiv \frac{1}{2c^{2}\left\vert (t^{\prime }-t)-\frac{1%
}{c^{2}}\frac{d\mathbf{r}(t^{\prime })}{dt^{\prime }}\cdot (\mathbf{r}%
^{\prime }\mathbf{-r})\right\vert }\right. \\
&&\frac{d}{dt^{\prime }}\left\{ v^{\prime \nu }(t^{\prime })\frac{R_{k}}{%
c^{2}\left\vert (t^{\prime }-t)-\frac{1}{c^{2}}\frac{d\mathbf{r}(t^{\prime })%
}{dt^{\prime }}\cdot (\mathbf{r-r}^{\prime })\right\vert }\right\} .  \notag
\end{eqnarray}

\emph{Proof} - In fact let us assume that the metric tensor $g_{\mu \nu }$
is constant and symmetric (Minkowski space-time). In this case the
synchronous variation of $\Delta S_{1}$ is given by Eqs.(\ref{Eq.---A.3})
and (\ref{Eq.---A.4}) where
\begin{eqnarray}
&&\left. d\left[ \int_{1}^{2}dr^{\prime \nu }\delta (R^{\alpha }R_{\alpha })%
\right] =dr^{k}\int_{t_{1}}^{t_{2}}cdt^{\prime }\right. \\
&&\sqrt{1-\frac{1}{c^{2}}\left\vert \frac{d\mathbf{r}^{\prime }}{dt^{\prime }%
}\right\vert ^{2}}u^{\prime \nu }(t^{\prime })\frac{\partial }{\partial r^{k}%
}\left[ \delta (R^{\alpha }R_{\alpha }-\sigma ^{2})\right] .  \notag
\end{eqnarray}%
Hence it follows,
\begin{eqnarray}
&&\left. \delta A=-4\left( \frac{q}{c}\right) ^{2}g_{\mu \nu }\frac{1}{4\pi }%
\int d\Sigma (\mathbf{n})\int_{1}^{2}\delta r^{\mu }dr^{k}\right.  \notag \\
&&\int_{t_{1}}^{t_{2}}cdt^{\prime }\sqrt{1-\frac{1}{c^{2}}\left\vert \frac{d%
\mathbf{r}^{\prime }}{dt^{\prime }}\right\vert ^{2}}u^{\prime \nu
}(t^{\prime }) \\
&&\frac{\partial }{\partial r^{k}}\left[ \delta (R^{\alpha }R_{\alpha
}-\sigma ^{2})\right] .  \notag
\end{eqnarray}%
while%
\begin{eqnarray}
&&\left. \delta B\equiv 4\left( \frac{q}{c}\right) ^{2}g_{\alpha \beta }%
\frac{1}{4\pi }\int d\Sigma (\mathbf{n})\int_{1}^{2}dr^{\alpha }\delta
r^{\mu }\right.  \notag \\
&&\int_{t_{1}}^{t_{2}}cdt^{\prime }\sqrt{1-\frac{1}{c^{2}}\left\vert \frac{d%
\mathbf{r}^{\prime }}{dt^{\prime }}\right\vert ^{2}}u^{\prime \beta
}(t^{\prime }) \\
&&\frac{\partial }{\partial r^{\mu }}\delta (R^{k}R_{k}-\sigma ^{2}).  \notag
\end{eqnarray}%
Let us now evaluate the partial derivative $\frac{\partial }{\partial r^{k}}%
\delta (R^{\alpha }R_{\alpha }-\sigma ^{2}).$ There results, thanks to the
chain rule
\begin{eqnarray}
&&\left. \frac{\partial }{\partial r^{k}}\delta (R^{\alpha }R_{\alpha
}-\sigma ^{2})=\right.  \label{ID-1} \\
&&\left. =\frac{\partial (R^{\alpha }R_{\alpha })}{\partial r^{k}}\frac{%
d\delta (R^{\alpha }R_{\alpha }-\sigma ^{2})}{d(R^{\alpha }R_{\alpha })}%
\equiv \right.  \notag \\
&&\left. \equiv 2R_{k}\frac{d\delta (R^{\alpha }R_{\alpha }-\sigma ^{2})}{%
dt^{\prime }}\frac{1}{\frac{d(R^{\alpha }R_{\alpha })}{dt^{\prime }}}\right.
.  \notag
\end{eqnarray}%
Hence, there follows the identity%
\begin{eqnarray}
&&\left. \frac{\partial }{\partial r^{k}}\delta (R^{\alpha }R_{\alpha
}-\sigma ^{2})=\right. \\
&&\left. =\frac{R_{k}}{c^{2}\left\vert (t-t^{\prime })-\frac{1}{c^{2}}\frac{d%
\mathbf{r}^{\prime }}{dt^{\prime }}\cdot (\mathbf{r-r}^{\prime })\right\vert
}\right.  \notag \\
&&\frac{d}{dt^{\prime }}\left\{ \delta (t-t^{\prime }-t_{ret})\right.  \notag
\\
&&\frac{1}{2c^{2}\left\vert (t-t^{\prime })-\frac{1}{c^{2}}\frac{d\mathbf{r}%
(t^{\prime })}{dt^{\prime }}\cdot (\mathbf{r-r}^{\prime })\right\vert },
\notag
\end{eqnarray}%
where $\mathbf{r}^{\prime }\equiv \mathbf{r}(t^{\prime }).$\ Integrating by
parts one obtains manifestly Eq.(\ref{Eq.---A.5-}). In a similar manner,
thanks again to Eq.(\ref{ID-1}), the term $\delta B$ can be cast in the form
(\ref{Eq.---A.6-}). \textbf{Q.E.D.}

\textbf{LEMMA 3 - Differential identity 1}

\emph{The following\ identity holds}

\begin{equation}
\frac{dt^{\prime }}{dt}=\frac{\sqrt{\sigma ^{2}+(\mathbf{r-r}^{\prime })^{2}}%
-\frac{d\mathbf{r}}{dt}\cdot (\mathbf{r-r}^{\prime })}{\sqrt{\sigma ^{2}+(%
\mathbf{r-r}^{\prime })^{2}}-\frac{d\mathbf{r}^{\prime }}{dt^{\prime }}\cdot
(\mathbf{r-r}^{\prime })}.
\end{equation}%
\emph{Proof} - In fact there results%
\begin{eqnarray}
&&\left. \frac{dt^{\prime }}{dt}=\frac{dt}{dt}-\frac{d}{dt}\sqrt{\sigma
^{2}+(\mathbf{r-r}^{\prime })^{2}}=1-\right. \\
&&-\frac{1}{\sqrt{\sigma ^{2}+(\mathbf{r-r}^{\prime })^{2}}}\left[ \frac{d%
\mathbf{r}}{dt}-\frac{dt^{\prime }}{dt}\frac{d\mathbf{r}^{\prime }}{%
dt^{\prime }}\right] \cdot (\mathbf{r-r}^{\prime })  \notag
\end{eqnarray}%
namely%
\begin{eqnarray}
&&\left. \frac{dt^{\prime }}{dt}\left[ 1-\frac{1}{\sqrt{\sigma ^{2}+(\mathbf{%
r-r}^{\prime })^{2}}}\frac{d\mathbf{r}^{\prime }}{dt^{\prime }}\cdot (%
\mathbf{r-r}^{\prime })\right] =\right. \\
&&\left. =1-\frac{1}{\sqrt{\sigma ^{2}+(\mathbf{r-r}^{\prime })^{2}}}\frac{d%
\mathbf{r}}{dt}\cdot (\mathbf{r-r}^{\prime }).\right.  \notag
\end{eqnarray}%
\textbf{Q.E.D.}

\section{Appendix E - Weakly relativistic approximation}

In validity of the asymptotic ordering (\ref{beta}) there results [from Eq.(%
\ref{Eq.-----30})] by Taylor expansion in $\beta ,$ while retaining exactly
all dependencies in terms of the retarded time $t^{\prime }$,

\begin{eqnarray}
&&\left. \overline{G}_{\mu }\cong \right.  \label{appen C-1} \\
&&\left. \cong 2c^{2}\left( \frac{q}{c}\right) ^{2}\left[ \frac{1}{%
c^{2}\left\vert (t-t^{\prime })\right\vert }\right. \right.  \notag \\
&&\left. \frac{d}{dt^{\prime }}\left\{ \frac{v_{\mu }^{\prime }(t^{\prime
})c(t-t^{\prime })-cR_{\mu }}{c^{2}\left\vert (t-t^{\prime })\right\vert }%
\right\} \right] _{t^{\prime }=t-t_{ret}}.  \notag
\end{eqnarray}

Instead, in the same approximation Eq.(\ref{Eq.--------30bis}) yields
\begin{eqnarray}
&&\overline{G}_{\mu }\cong  \label{appen C-2} \\
&&\left. \cong 2c\left( \frac{q}{c}\right) ^{2}\frac{1}{c^{2}\left\vert
(t-t^{\prime })\right\vert ^{2}}\left[ \frac{dr_{\mu }(t-t_{ret})}{dt}%
+\right. \right.  \notag \\
&&\left. -R_{\mu }^{\prime }\frac{c^{2}}{c^{2}\left\vert (t-t^{\prime
})\right\vert }\right]  \notag
\end{eqnarray}%
where\emph{\ }$R^{\prime \alpha }\equiv \left\{ ct_{ret},\mathbf{r}(t)%
\mathbf{-r}(t-t_{ret})\right\} .$\emph{\ }Here the delay time $t_{ret}\equiv
t-t^{\prime },$ evaluated in a similar way from Eq.(\ref{Eq.-alfa-3})
neglecting corrections of order $\beta $, reads$:$
\begin{equation}
t_{ret}\cong \frac{\sigma }{c}.  \label{App-C-1}
\end{equation}%
It follows%
\begin{equation}
\overline{G}_{\mu }\cong (0,\mathbf{G}).
\end{equation}%
In particular, the spatial 3-vector $\mathbf{G}$ reads in case of Eq.(\ref%
{appen C-1}) :%
\begin{eqnarray}
&&\left. \mathbf{G}\cong \right. \\
&&\left. \cong -2c\left( \frac{q}{c}\right) ^{2}\left[ \frac{1}{%
c^{2}\left\vert (t-t^{\prime })\right\vert }\right. \right.  \notag \\
&&\left. \left. \frac{d}{dt^{\prime }}\left\{ \mathbf{v}(t^{\prime })-\frac{%
\mathbf{r}(t)\mathbf{-r}(t^{\prime })}{(t-t^{\prime })}\right\} \right]
_{t^{\prime }=t-t_{ret}}\right. .  \notag
\end{eqnarray}%
This equations, with (\ref{App-C-1}), implies Eq.(\ref{Eq.50}). Finally, let
us evaluate also the corresponding short-time approximation, obtained
invoking also the ordering (\ref{ordering -1}). By Taylor expansion in \ $%
\xi \equiv (t-t^{\prime })/t$ there results to leading order

\begin{equation}
\left[ \frac{d}{dt^{\prime }}\mathbf{v}(t^{\prime })+\frac{\mathbf{v(}%
t^{\prime })}{(t-t^{\prime })}-\frac{\mathbf{r}(t)\mathbf{-r}(t^{\prime })}{%
(t-t^{\prime })^{2}}\right] _{t^{\prime }=t-t_{ret}}\cong
\end{equation}%
\begin{equation*}
\cong \frac{1}{2}\frac{d}{dt}\mathbf{v(}t)-\frac{R}{3c}\frac{d^{2}}{dt^{2}}%
\mathbf{v(}t).
\end{equation*}%
Therefore, one obtains finally the weakly-relativistic (and short-time)
approximation
\begin{equation}
\mathbf{G\cong }\left( \frac{q}{c}\right) ^{2}\left[ -\frac{1}{\sigma }\frac{%
d}{dt}\mathbf{v(}t)+\frac{2}{3c}\frac{d^{2}}{dt^{2}}\mathbf{v(}t)\right] ,
\label{appen  C-3}
\end{equation}%
which similarly recovers\ Eq.(\ref{weakly-relativistic RR force}). \
Instead, in the case of Eq.(\ref{appen C-2}) in an analogous way there
results:%
\begin{eqnarray}
&&\mathbf{G}\cong \\
&&\left. \cong 2c\left( \frac{q}{c}\right) ^{2}\frac{1}{c^{2}\left\vert
(t-t^{\prime })\right\vert ^{2}}\left[ \frac{d\mathbf{r}(t-t_{ret})}{dt}%
-\right. \right.  \notag \\
&&\left. -\frac{\mathbf{r}(t)\mathbf{-r}(t-t_{ret})}{\left\vert (t-t^{\prime
})\right\vert }\right] ,  \notag
\end{eqnarray}%
which implies Eq.(\ref{Eq.51}). Hence it follows%
\begin{eqnarray}
&&\left. \frac{d\mathbf{r}(t-t_{ret})}{dt}-\frac{\mathbf{r}(t)\mathbf{-r}%
(t-t_{ret})}{\left\vert (t-t^{\prime })\right\vert }\cong \right. \\
&&\left. \cong -\frac{R}{2c}\frac{d^{2}\mathbf{r}(t)}{dt^{2}}+\frac{1}{3}%
\frac{R^{2}}{2c^{2}}\frac{d^{3}\mathbf{r}(t)}{dt^{3}},\right.  \notag
\end{eqnarray}%
which implies again Eq.(\ref{appen C-3}) and therefore recovers the same
weakly-relativistic approximation given by Eq. (\ref{weakly-relativistic RR
force}).

\end{document}